\documentclass[12pt]{article}
\usepackage{amssymb}
\usepackage{amsmath}
\usepackage{amsfonts}
\usepackage{hyperref}
\usepackage{color, xcolor}
\usepackage[T1]{fontenc}

\usepackage{mathrsfs}

\definecolor{blue-violet}{rgb}{0.54, 0.17, 0.89}
\definecolor{PineGreen}{cmyk}{0.92, 0, 0.59, 0.25}
\definecolor{OliveGreen}{cmyk}{0.64, 0, 0.95, 0.40}
\definecolor{RawSienna}{cmyk}{0, 0.72, 1, 0.45}
\definecolor{Gray}{cmyk}{0, 0, 0, 0.50}
\definecolor{MidnightBlue}{cmyk}{0.98, 0.13, 0, 0.43}
\definecolor{Orange}{cmyk}{0, 0.61, 0.87, 0}
\definecolor{LimeGreen}{cmyk}{0.50, 0, 1, 0}
\definecolor{Green}{cmyk}{1, 0, 1, 0}
\definecolor{JazzberryJam}{rgb}{0.65, 0.04, 0.37}

\newcommand{\Diff}{\mathrm{D}}
\newcommand{\diff}{\mathrm{d}}

\usepackage[makeroom]{cancel}
\usepackage[normalem]{ulem}

\numberwithin{equation}{section}

\begin{document}

\title{\bf Thermodynamics of Chern-Simons AdS$_5$ black holes coupled to $\mathrm{SU}(2)$ solitons}

\author{Laura Andrianopoli $^{1,}$\thanks{ laura.andrianopoli@polito.it}\;,  
Du\v san \DJ or\dj evi\'c $^{2,}$\thanks{dusan.djordjevic@ff.bg.ac.rs } \\
and Olivera Miskovic $^{3,1,}$\thanks{olivera.miskovic@pucv.cl} 
\bigskip\\
{\small\it $^1$DISAT, Politecnico di Torino, Corso Duca degli Abruzzi, 24, 10129 Torino, Italy and}\\
{\small\it  INFN, Sezione di Torino, via P. Giuria 1, 10125 Torino, Italy.} \\
{\small\it $^2$Faculty of Physics, University of Belgrade, Studentski Trg 12-16, 11000 Belgrade, Serbia.} \\
{\small\it $^3$Instituto de F\'\i sica, Pontificia Universidad Cat\'olica de Valpara\'\i so,}\\ 
{\small\it Avda.~Universidad 330, Curauma, Valpara\'{\i}so, Chile.}}

\maketitle

\begin{abstract} 
We investigate properties of the five-dimensional Chern--Simons AdS
black hole coupled to $\mathrm{SU}(2)$ solitons by means of a minisuperspace
approximation adapted to static, spherically symmetric configurations. The
reduced action reproduces the known branch of solutions and provides a
variational framework in which the boundary terms determine the conserved
quantities and their conjugate variables. In particular, we recover the
energy and the $\mathrm{U}(1)$ charge previously obtained by Hamiltonian
methods, while the enlarged parameter space also reveals a momentum
conjugate to the trace-torsion mode. The Euclidean action yields an entropy
satisfying the first law of black hole thermodynamics. In contrast to many
other torsional black hole models, the axial torsion parameter, describing the secondary black hole hair, together with the trace-torsion mode,
contributes nontrivially to the entropy. The expression for the entropy obtained in this way is further confirmed by the other two methods found in the literature. 

\end{abstract}

\tableofcontents

%%%%%%%%%%%%%%%%%%%%%%%%%%%%%%%%%%%%%%%%%%
\section{Introduction}
%%%%%%%%%%%%%%%%%%%%%%%%%%%%%%%%%%%%%%%%%%

Chern--Simons (CS) anti-de Sitter (AdS) supergravity \cite{Chamseddine:1976bf,Troncoso:1998ng} (see also \cite{Zanelli:2005sa}) provides a distinguished class of gravitational theories in odd spacetime dimensions
whose structure differs substantially from that of General Relativity. In contrast to Einstein gravity, it is formulated as a genuine gauge theory for the AdS group, so that both gravity and supergravity arise from a gauge principle rather than from the metric alone. For this reason, CS AdS gravity
is not intended to describe a realistic gravitational interaction in the usual phenomenological sense. Nevertheless, it offers a valuable theoretical framework in which several structural aspects of gravitational physics can be studied in a particularly sharp way. In three dimensions, it reproduces
gravity as a topological theory, while in higher odd dimensions it ceases to be topological and develops local dynamics \cite{Banados:1996yj}, and can even possess more than one different 
dynamical sectors in the phase space \cite{Miskovic:2003ex}. Moreover, it admits black hole solutions \cite{Banados:1993ur,Aros:2002rk,Zegers:2005vx,Deser:2005gr,Canfora:2007xs,Brihaye:2013vsa,Giribet:2014hpa,Andrianopoli:2021qli}, whose asymptotic behavior differs from that of the Schwarzschild--AdS family, as well as extremal supersymmetric configurations
in suitable supergravity extensions \cite{Miskovic:2006ei,Canfora:2007xs,Edelstein:2010sx,Andrianopoli:2021qli}. These properties make CS AdS gravity a natural laboratory in which to investigate which features of black hole physics are specific to Einstein gravity and which ones are more universal.

A particularly interesting question concerns black hole thermodynamics.
Since CS gravity differs from General Relativity both in its gauge-theoretic
origin and in the role played by torsion, it is not a priori clear to what
extent the familiar thermodynamic structure survives. Indeed, standard
relations that hold in Einstein-AdS gravity, such as the usual form of the
Smarr formula, do not straightforwardly extend to the CS framework. For instance, in Lovelock gravity, the Smarr relation acquires additional work terms \cite{Liberati:2015xcp}. It is
therefore natural to ask whether the conserved charges, entropy, and first
law of black hole thermodynamics can still be defined consistently, and how
they are modified by the presence of additional fields and nontrivial
torsional degrees of freedom. From this perspective, thermodynamics provides
a useful probe of the internal consistency of the theory and of the physical
role of its non-Riemannian sector.

Black hole thermodynamics in the presence of torsion has been explored in
several contexts. Part of this interest comes from first-order formulations
of gravity, where torsion plays a natural geometric role, and also from
theories such as teleparallel gravity, in which torsion rather than
curvature is the fundamental field strength off shell. In three dimensions, where the analysis is technically simpler and CS coincides with Einstein gravity, some results are already known.  In particular, the thermodynamics of the BTZ black hole has been
studied in models with torsional degrees of freedom, and both the conserved
charges and the first law have been established in Hamiltonian and
Lagrangian approaches \cite{Blagojevic:2006jk,Blagojevic:2006nf}. In these
examples, the presence of a gravitational CS term modifies the black hole
entropy by an additional contribution associated with torsion, see also \cite{Ma:2013eaa}. The same system was later revisited from a holographic
perspective, where the boundary stress tensor and conformal structure
reproduce the same thermodynamic quantities \cite{Klemm:2007yu}, and more
recent generalizations to enlarged symmetry algebras have also been
considered \cite{Aviles:2023igk}. More generally, a first-order treatment of
black hole entropy and the first law on Riemann--Cartan spacetimes was
developed in \cite{Blagojevic:2019gsd} and applied to explicit solutions in 
\cite{Blagojevic:2022etm}. It has also been argued, under specific
assumptions, that torsion need not contribute explicitly to the black hole
entropy \cite{Chakraborty:2018qew}. In five dimensions, black holes with
torsion have been studied in Lovelock theories as well \cite{Cvetkovic:2017nkg}, but the role of torsion in the thermodynamic
description, there, remains rather different from the one found in CS gravity. More
broadly, the first law has also been analyzed in higher-dimensional theories
coupled to gauge fields and CS terms by means of Wald's covariant formalism 
\cite{Rogatko:2006hck}, and to scalar fields in Lovelock gravity in \cite{
Correa:2013bza,Bravo-Gaete:2021hza}.

A further motivation comes from the AdS/CFT correspondence and holographic applications \cite{Maldacena:1997re,Witten:1998qj,GubserAdSCFT2002},
investigated in the framework of CS AdS gravity in  \cite{Banados:2005rz,Cvetkovic:2017fxa}.  The existence of asymptotically AdS black holes in CS AdS gravity, with an associated Hawking temperature, suggests that the corresponding dual field
theories should describe thermal states. From this viewpoint, the thermodynamics of CS AdS black holes may provide information about a class of (possibly realistic) strongly coupled thermal systems that differ from the ones holographically related to the standard Einstein-gravity setting. In addition, five-dimensional CS gravity is one of the few theories in which the holographic role of torsion is understood in some detail, as a source of the fermionic spin currents on the boundary \cite{Banados:2006fe}. In particular, it has been used as a toy model for holographic systems with spin current \cite{Gallegos:2020otk}, and in \cite{Juricic:2024tbe} as a model
of the strongly coupled system with dislocations. This makes the study of
black hole thermodynamics especially interesting in a framework where
torsion, non-Abelian structure, and holography are simultaneously present.
Since torsion has also been related holographically to defects or
dislocations in the boundary theory, the analysis of thermal black hole
configurations may help clarify how these non-Riemannian bulk features are
encoded in the dual description.

In this paper, we study the thermodynamics of a five-dimensional CS AdS black hole within the minisuperspace
approximation. The solution under consideration is characterized by the
presence of two non-Abelian solitons, identified by nontrivial Pontryagin
numbers, one of which (axial torsion) gives rise to secondary hair. Our aim
is to determine how these features are incorporated into a consistent
thermodynamic description. To this end, we work with the Euclidean reduced
action, which provides a convenient framework for implementing the action
principle and extracting the relevant thermodynamic information. Within this
approach, we identify the conserved charges of the solution, compute the
black hole entropy, and show that the resulting quantities satisfy the first
law of thermodynamics.

Our results reveal that the thermodynamic structure of these solutions
differs significantly from the Einstein-gravity pattern. In particular, the
entropy does not obey the standard area law, but depends nontrivially on
additional integration constants. This shows that, in the present CS
setting, the non-Abelian sector and the torsional structure affect the black
hole thermodynamics in a genuinely nontrivial way. The minisuperspace
approximation is especially useful in this respect, since it retains a
sufficiently rich sector of the theory to capture these effects while
keeping the analysis under analytic control. At the same time, the Euclidean
formalism makes it possible to relate the finiteness and stationarity of the
action to the definition of entropy and to the validity of the first law.

The paper is organized as follows. In Sec.~\ref{review}, we review the
relevant features of CS AdS supergravity and present the class of black hole
configurations considered in this work. In Sec.~\ref{mini}, we introduce the
minisuperspace approximation and construct the Euclidean reduced action. In
Sec.~\ref{lorentz}, we analyze the action principle in Lorentzian signature
and derive the conserved quantities, showing that they coincide with those
obtained by Hamiltonian methods. In Sec.~\ref{euclid}, we perform the
Euclidean continuation of the spacetime and compute the entropy. We then
show that the resulting quantities satisfy the first law of black hole
thermodynamics. In Sec.~\ref{other}, we compare our expression for the
entropy with two other formulas proposed in the literature for
non-Riemannian spacetimes. Finally, in Sec.~\ref{conclusions}, we discuss
the implications of our results for the role of non-Abelian hair, torsion,
and the possible universality of black hole thermodynamics in CS AdS gravity.

%%%%%%%%%%%%%%%%%%%%%%%%%%%%%%%%%%%%%%%%%%
\section{Chern-Simons AdS black hole}
\label{review}
%%%%%%%%%%%%%%%%%%%%%%%%%%%%%%%%%%%%%%%%%%

Consider the five-dimensional CS AdS supergravity \cite{Chamseddine:1976bf}, invariant under $\mathrm{SU}(2,2|\mathcal{N})$ gauge group. The bosonic sector $\mathrm{SU}(2,2)  \times\mathrm{SU} (\mathcal{N}) \times \mathrm{U}(1)$ of the gauge field has 1-form  components: the vielbein $e^{a}$ and the spin connection $\omega ^{ab}$, associated with the AdS$_5$ isometry algebra $\mathrm{SU}(2,2)\simeq \mathrm{SO}(2,4)$, the  $\mathrm{U}(1)$ gauge field $A$ and non-Abelian $\mathrm{SU}(\mathcal{N})$ gauge field $\mathcal{A}$.  The associated field strength 2-forms are: the AdS ones, that is the curvature $R^{ab}=\mathrm{d}\omega ^{ab}+\omega
^{ac}\omega _{c}^{\ b}$ and the torsion $T^{a}=\mathrm{D}(\omega)e^{a}$, the Abelian field strength $F=\mathrm{d}A$ and the  $\mathrm{SU}(\mathcal{N})$ field strength $\mathcal{F}=\mathrm{d}\mathcal{A}+ \mathcal{A}^{2}$.  
We omitted writing the wedge product between the forms. 

The case $\mathcal{N}=4$ is special because the Abelian kinetic term vanishes, making the gauge field $A$ non-dynamical. As a consequence,  the theory exhibits distinctive topological properties and admits nontrivial solutions \cite{Miskovic:2006ei,Edelstein:2010sx,Giribet:2014hpa,Andrianopoli:2021qli}.

The black hole solution of interest is found in the $\mathcal{N}=4$  CS AdS supergravity. The bosonic action is given by \cite{Chamseddine:1976bf}
\begin{equation}
    \mathring{I}[\mathbf{A}]=k\int \left[ \rule[2pt]{0pt}{11pt}L_{\mathrm{G}}(e,\omega )+L_{\mathrm{SU}(4)}(\mathcal{A})+L_{\mathrm{int}}(A,e,\omega , \mathcal{A})\right] \,,  \label{action}
\end{equation}
where $k$ is the gravitational constant and the level of the CS form and, up to boundary terms,  
\begin{eqnarray}
L_{\mathrm{G}} &=&\frac{1}{8\ell }\,\epsilon _{abcde}\left(
R^{ab}R^{cd}e^{e}+\frac{2}{3\ell ^{2}}\,R^{ab}e^{c}e^{d}e^{e}+\frac{1}{5\ell
^{4}}\,e^{a}e^{b}e^{c}e^{d}e^{e}\right) \,,  \notag \\
L_{\mathrm{SU}(4)} &=&-\frac{\mathrm{i}}{3}\,\mathrm{Tr}\left( \mathcal{AF}%
^{2}-\frac{1}{2}\,\mathcal{A}^{3}\mathcal{F}+\frac{1}{10}\,\mathcal{A}%
^{5}\right) \,,  \notag \\
L_{\mathrm{int}} &=&\frac{1}{4}\left( \frac{1}{2}\,R^{ab}R_{ab}+\frac{1}{\ell ^{2}}\,R^{ab}e_{a}e_{b}-\frac{1}{\ell ^{2}}\,T^{a}T_{a}-
\mathcal{F}^{I}\mathcal{F}_{I}\right) \,A\,,  \label{L}
\end{eqnarray}
while $L_{\mathrm{U}(1)}=0$ when $\mathcal{N}=4$. In the above expressions, the Lorentz indices are denoted by $a=0,1,\cdots, 4$, while the $\mathrm{SU}(4)$ adjoint indices are denoted by $I$. The action has only two coupling constants: the level $k$ related to the five-dimensional analog of the Newton constant $G_{\mathrm{N}}=\ell ^{3}/8\pi k$, and the AdS radius $\ell $ with the negative cosmological constant normalized as\footnote{\label{radius} 
The AdS radius in CS gravity, $\ell$, differs from the AdS radius in Einstein-Hilbert gravity, $\ell_0 =\ell \sqrt{2}$, associated with the standard form of the cosmological constant in five dimensions, $\Lambda=-6/\ell_0^2$.} $\Lambda =-3/\ell ^{2}$. The supersymmetry fixes all other interactions, as is typical for CS AdS supergravities in higher dimensions \cite{Troncoso:1998ng}.

From \eqref{L}, the gravitational Lagrangian $L_{\mathrm{G}}$ is the Einstein-Gauss-Bonnet action in the CS point, that is, with the Gauss-Bonnet coupling fixed as $\alpha _{\mathrm{CS}}=\ell ^{2}/4$. 
Furthermore, $L_{\mathrm{SU}(4)}$ is the CS Lagrangian for $\mathrm{SU}(4)$ gauge group, and the interaction term $L_{\mathrm{int}}$ couples the gauge field $A$ with the Pontryagin invariants for Lorentz and $\mathrm{SU}(4)$ groups, respectively. 
The difference in the relative sign between the AdS sector $\frac{1}{2}\,R^{ab}R_{ab}+\frac{1}{\ell ^{2}}\,R^{ab}e_{a}e_{b}-\frac{1}{\ell ^{2}}\,T^{a}T_{a}$ and the internal sector $\mathcal{F}^{I}\mathcal{F}_{I}$ is due to the supertrace in the supergroup. 

%\gr{The gravitational Lagrangian $L_{\mathrm{G}}$ is the Einstein-Gauss-Bonnet action in the CS point, that is, with the five-dimensional analog of the Newton constant $G_{\mathrm{N}}=\ell ^{3}/8\pi k$, where $\ell $ is the AdS radius, negative cosmological constant normalized as $\Lambda =-3/\ell ^{2}$, and the Gauss-Bonnet coupling fixed as $\alpha _{\mathrm{CS}}=\ell ^{2}/4$.  Furthermore, $L_{\mathrm{SU}(4)}$ is the CS Lagrangian for $\mathrm{SU}(4)$ gauge group, and the interaction term $L_{\mathrm{int}}$ couples the gauge field $A$ with the Pontryagin invariants for Lorentz and $\mathrm{SU}(4)$ groups, respectively. The difference in the relative sign between the AdS sector $\frac{1}{2}\,R^{ab}R_{ab}$ and the internal sector $\mathcal{F}^{I}\mathcal{F}_{I}$ is due to the supertrace in the supergroup. The only arbitrary coupling constant is the gravitational one, and the supersymmetry fixes all other interactions.}

The least action principle leads to equations of motion whose bosonic part reads
\begin{eqnarray}
    \mathcal{E}_{a} &=&\frac{1}{8}\,\epsilon _{abcde}\,\left( R^{bc}R^{de}+\frac{2}{\ell ^{2}}\,R^{bc}e^{d}e^{e}+\frac{1}{\ell ^{4}}\,e^{b}e^{c}e^{d}e^{e}\right) -\frac{1}{2\ell }\,T_{a}F\,,  \notag \\
    \mathcal{E}_{ab} &=&\frac{1}{2\ell }\,\epsilon _{abcde}\,\left( R^{cd}+\frac{1}{\ell ^{2}}\,e^{c}e^{d}\right) T^{e}+\frac{1}{2}\,\left( R_{ab}+\frac{1}{\ell ^{2}}\,e_{a}e_{b}\right) F\,,  \notag \\
    \mathcal{E} &=&\frac{1}{8}\,\left( R^{ab}R_{ab}+\frac{2}{\ell ^{2}} \,R^{ab}e_{a}e_{b}-\frac{2}{\ell ^{2}}\,T^{a}T_{a}\right) -\frac{1}{4}\,\mathcal{F}^{I}\mathcal{F}_{I}\,,  \notag \\
    \mathcal{E}_{I} &=&-\frac{1}{2}\,\mathcal{F}_{I}F+d_{II_{1}I_{2}}\mathcal{F}^{I_{1}}\mathcal{F}^{I_{2}}\,,  \label{EOM}
\end{eqnarray}
$d_{I_{1}I_{2}I_{3}}$ being the symmetric invariant tensor of $\mathrm{SU}(4)$. As mentioned before, the $\mathrm{U}(1)$ gauge field does not have a kinetic term in the equations of motion because it is non-dynamical. One particular solution of \eqref{EOM} is the global AdS$_{5}$ space with constant negative curvature, $R^{ab}=-\frac{1}{\ell ^{2}}\,e^{a}e^{b}$, and all other field strengths equal to zero.

The non-Abelian sector significantly simplifies when the internal symmetry is
broken as $\mathrm{SU}(4)\rightarrow \mathrm{SU}(2)$.\footnote{In the original work \cite{Andrianopoli:2021qli}, the non-Abelian internal
symmetry $\mathrm{SU}(4)$ was broken as $\mathrm{SU}(4)\rightarrow \mathrm{SU}(2)\times \mathrm{U}(1)_{c}$, where the subscript  `$c$' was introduced to distinguish this new Abelian subgroup
from the original $\mathrm{U}(1)$ factor present in the bosonic sector of $\mathrm{SU}(2,2|\mathcal{N})$. Although the corresponding $\mathrm{U}(1)_{c}$ gauge field appears as a pure gauge configuration, it played a key role in constructing a solution carrying a nontrivial $\mathrm{U}(1)_{c}$ charge. In
the present context, however, the $\mathrm{U}(1)_{c}$ symmetry does not play
a significant role and will therefore be omitted from our analysis.} Then
the symmetric rank-3 tensor is identically zero, $d_{I_{1}I_{2}I_{3}}\equiv 0
$, and the kinetic term of the $\mathrm{SU}(2)$ gauge field vanishes, $L_{\mathrm{SU}(4)\rightarrow \mathrm{SU}(2)}=0$, making the $\mathrm{SU}(2)$ gauge field non-dynamical. The only dynamical fields in the theory are the gravitational ones, related to the metric and torsion.

Choosing the spherical coordinates as $x^{\mu }=(t,r,y^{m})$, where $y^{m}$ ($m=2,3,4$) are three angles, the corresponding coordinates on the tangent bundle are labeled by $a=(0,1,i)$, with $i=2,3,4$. Then we can write the following solution of the equations \eqref{EOM}  with $d_{I_{1}I_{2}I_{3}}= 0
$ \cite{Andrianopoli:2021qli},  
\begin{equation}
\begin{array}[b]{llllll}
    e^{0} & =\hat{N}\hat{f}\,\mathrm{d}t\,,\medskip  & e^{1} & =\dfrac{\mathrm{d}r}{\hat f}\,, & e^{i} & =r\tilde{e}^{i}\,, \\
    \omega ^{01} & =
    \hat f(\hat N\hat f)'\,\mathrm{d}t\,,\medskip  & \omega ^{0i} & =0\,, & \omega ^{1i} & =-\hat f\,\tilde{e}^{i}\,, \\
    \omega ^{i} & =\tilde{\omega}^{i}-C\,\tilde{e}^{i}\,,\qquad  & \mathcal{A}^{I} & =\delta^I_i\left(\tilde{\omega}^{i}-B\,\tilde{e}^{i}\right)\,,\qquad  & A & = \mathrm{d}\lambda \,.
\end{array}
\label{solution}
\end{equation}
 The 1-form $\omega^i = \frac 12 \,\epsilon^i{}_{jk}\,\omega^{jk}$ provides a convenient representation of the transverse components of the spin connection, with the convention $\epsilon_{trijk}=\epsilon_{ijk}$. The quantity $\tilde\omega^i = \tilde\omega^i (\tilde e)$ denotes its torsionless part, which also plays the role of the internal (torsionless) connection on the unit 3-sphere corresponding to the transversal section of the spacetime. 
Owing to the isomorphism between the tangent space subgroup $\mathrm{SU}(2)_{\mathrm{Lorentz}} \subset \mathrm{SO}(2,4)$ and the internal subgroup $\mathrm{SU}(2) \subset \mathrm{SU}(4)$, the corresponding group indices can be identified for the solitonic configurations considered here, namely $i=I$. This on-shell identification allows the solution to possess unbroken supersymmetries because it permits the construction of Killing spinors for the solution \eqref{EOM} in the BPS limit by ensuring that the associated generators act identically on the Killing spinors \cite{Andrianopoli:2021qli}. 

The metric obtained from \eqref{solution} describes a static, spherically symmetric black hole geometry,
\begin{equation}
    \mathrm{d}s^{2}=\eta _{ab}\,e^{a}\otimes e^{b}=-\hat N^{2}\hat f^{2}\,\mathrm{d}t^{2}+\frac{\mathrm{d}r^{2}}{\hat f^{2}}+r^{2}\mathrm{d}\Omega ^{2}\,,  \label{metric}
\end{equation}
with the transversal 3-sphere section $\mathrm{d}\Omega ^{2}=\gamma
_{mn}(y)\,\mathrm{d}y^{m}\mathrm{d}y^{n}$, and $\gamma _{mn}=\eta _{ij}\,
\tilde{e}_{m}^{i}\tilde{e}_{n}^{j}$. The metric functions have the form\footnote{It is possible to generalize this solution by introducing the gravitational hair $(b,b_0)$ that changes the asymptotic structure of spacetime to $\hat f^2 \sim \frac{r^2}{\ell^2}+br+1-\mu$, and similarly for $\hat f\hat N$ when $b$ is replaced by $b_0$ \cite{Andrianopoli:2021qli}. However, they also introduce new singularities, some of them on the horizon, so we will not consider them in our study.}
\begin{equation}
    \hat N=N_{0}=1\,,\qquad \hat f=\sqrt{\frac{r^{2}}{\ell ^{2}}-e^{2}}\,\qquad e^{2}=\mu-1\geq 0\,.  \label{N,f}
\end{equation}
The extremality parameter $e^{2}\geq 0$ depends on the mass parameter $\mu $ (integration constant), and the integration constant $N_{0}=1$ normalized such that the space is asymptotically AdS in the Riemann sector. The non-negativity of $e$ ensures the existence of the horizon. The horizon $r_{+}$ can be related to other parameters in the following way,
\begin{equation}
    e=\frac{r_{+}}{\ell }\,,\qquad r_{+}=\ell \sqrt{\mu -1}\,,\qquad e^{2}=\mu-1\,.
\end{equation}
The Hawking temperature of the black hole is linear in the horizon radius,
\begin{equation}
    T=\frac{\hat N(r_{+})(\hat f^{2})'_{r_{+}}}{4\pi }=\frac{N_{0}r_{+}}{2\pi\ell ^{2}}=\frac{N_{0}e}{2\pi \ell }\,,\qquad N_{0}=1\,.  
    \label{T}
\end{equation}
On the other hand, the $\mathrm{SU}(2)$ field describes a soliton  
winding around the 3-sphere, of magnitude $B$, where $B$ is the third integration constant. Another $\mathrm{SU}(2)$ soliton, of magnitude $C$, is realized by the axial torsion field. Finally, the $\mathrm{U}(1)$ field is pure gauge in the spherically symmetric ansatz, as expected from a non-dynamical field. In total, there are four integration constants $(\mu,N_{0},C,B)$. The supersymmetric BPS state is realized in the extremal limit $\mu=1$, $N_0=1$ and $B=C$ \cite{Andrianopoli:2021qli}.

The above configuration can be better understood geometrically if we write down the corresponding non-vanishing field strength components,
\begin{equation}
\begin{array}[b]{llll}
T^{i} & =rC\,\epsilon _{\ jk}^{i}\,\tilde{e}^{j}\tilde{e}^{k}\,,\medskip  & 
R^{1i} & =-\dfrac{1}{\ell ^{2}}\,e^{1}e^{i}-C\hat{f}\,\epsilon _{jk}^{\ \ i}
\tilde{e}^{j}\tilde{e}^{k}\,, \\ 
\mathcal{F}^{i} & =\dfrac{\tilde{\Xi}_{0}}{2}\,\epsilon _{\ jk}^{i}\,\tilde{e}^{j}\tilde{e}^{k}\,,\medskip \qquad  & R^{ij} & =\left( \Xi _{0}-\dfrac{r^{2}}{\ell ^{2}}\right) \,\epsilon _{\ jk}^{i}\, \tilde{e}^{j}\tilde{e}^{k}\,.
\end{array}
\end{equation}
Since $F^{ab}=R^{ab}+\frac{1}{\ell ^{2}}\,e^{a}e^{b}$ is the Lorentz component of the AdS field strength, its restriction to the transversal 3-sphere $\mathbb{S}^3\simeq \mathrm{SU}(2)$,
\begin{equation}
F^{i}=\frac{1}{2}\,\epsilon _{\ jk}^{i}\,F^{jk}=\dfrac{\Xi _{0}}{2}
\,\epsilon _{\ jk}^{i}\, \tilde{e}^{j}\tilde{e}^{k}\,, 
\label{F}
\end{equation}
takes the same form as the field strength of the internal $\mathrm{SU}(2)$ group, differing only in its magnitude. This matching allows for the construction of BPS states in the theory \cite{Andrianopoli:2021qli}.

The field is asymptotically AdS in the Riemannian sector, but the full field strength does not vanish on the
boundary. Namely, the 3-sphere curvature \eqref{F} has the strength $\Xi_0\neq 0$,
while the internal non-Abelian field has the strength $\tilde\Xi_0\neq 0$, and they are related to the integration constants as
\begin{equation}
    \Xi_0=e^{2}+1-C^{2}\,,\qquad \tilde\Xi_0=1-B^{2}\leq 1\,. \label{Xi-charge}
\end{equation}

The obtained solution carries both topological and Noether charges. The topological charges are the $\mathrm{SU}(2)$-Pontryagin index of each soliton,
\begin{equation}
  \label{n12}  n_1=-C\left( 2-C^{2}\right) \in\mathbb{Z}\,,\qquad n_2=B(2-B^{2})\in \mathbb{Z} \,,
\end{equation}
describing how many times each of them winds around the 3-sphere. On the other hand, the Noether charges were computed using the Regge-Teitelboim
method \cite{Regge-Teitelboim} in the Hamiltonian formalism, which determines only their variations. For the  $\mathrm{U}(1)$ charge, it is found 
\begin{equation}
    \delta \hat Q=-3k\pi ^{2}\left[ \rule{0pt}{13pt}\delta (Ce^{2})+(1-C^{2})\, \delta C-(1-B^{2})\,\delta B\right] \,,
\end{equation}
which, when integrated, leads to a conserved quantity not independent from the topological charges,
\begin{equation}
    \hat Q =\, 3k\pi ^{2}\left( -C+\frac{1}{3}C^{3}-Ce^{2}-\frac{1}{3}B^{3}+B\right)  \,.
    \label{hatQ}
\end{equation}
The integration constant is chosen so that $\hat Q =0$ without solitons (that is, when $C=0$, $B=0$).
As for the energy, on the other hand, its variation is given by 
\begin{equation}
    \delta \hat E=\frac{3k\pi ^{2}}{\ell }\,\left[ \rule{0pt}{13pt}\left(e^{2}+1-C^{2}\right) \,\delta e^{2}-2e^{2}\delta C^{2}\right]. \label{hatE}
\end{equation}
It is not integrable in general, but it can be solved in our case because the soliton charge is quantized, thus $\delta C=0$. Namely, eq.~\eqref{n12} gives $\delta n_1=0$, which implies $\delta C=0$. Then, the total energy reads
\begin{equation}
    \hat E=\frac{3k\pi ^{2}}{\ell }\,\left( \frac{e^{2}}{2}+1-C^{2}\right) \,e^{2}\,.
\end{equation}
Our goal is to investigate the thermodynamics of the obtained solution, which can be done in a suitable approximation.

%%%%%%%%%%%%%%%%%%%%%%%%%%%%%%%%%%%%%%%%%%
\section{The minisuperspace  approximation}\label{mini}
%%%%%%%%%%%%%%%%%%%%%%%%%%%%%%%%%%%%%%%%%%

The minisuperspace approximation is a well-established method that simplifies the analysis by retaining only those degrees of freedom relevant to a specific class of symmetric configurations. In our case, it isolates the modes compatible with spherical symmetry and staticity, while all others are consistently suppressed. This reduction is particularly useful for studying the radial dynamics of gravitational systems, as the symmetry constraints imply that all fields depend solely on the radial coordinate $r>0$, which effectively plays the role of an evolution parameter. The resulting dynamics are then governed by the Hamiltonian formulation of the reduced (minisupersymmetric) action, following the approach developed in Refs.~\cite{Crisostomo:2000bb,Aros:2000ij}.\footnote{\label{BH scan} Comparison with the article `Black Hole Scan' \cite{Crisostomo:2000bb} is done based on eqs.~(10) and (14) of that paper, by setting the dimension $d=2n-1=5$, the curvature polynomial degree $k=2$, the unit 3-sphere volume $\Omega^{\mathrm{BHscan}} _{3}=\mathrm{Vol}(\Sigma )=2\pi ^{2}$,
and the gravitational coupling constant $G_{2}=\frac{2\ell }{3k\mathrm{Vol}(\Sigma )}=
\frac{\ell }{3k\pi ^{2}}$. Notice that the volume convention in that article $\Omega^{\mathrm{BHscan}} _{3}$ does not coincide with the volume convention $\Omega_{3}=6k\mathrm{Vol}(\Sigma )$ used in this manuscript.}

The consistency of the approximation requires that applying the minisuperspace ansatz directly to the action yields equations of motion equivalent to those obtained by first varying the full action and then imposing the same ansatz. In other words, the variational principle applied to the reduced action must produce the correct dynamics for the symmetric sector. This consistency is
guaranteed under the conditions established by the reduction theorems of  Palais \cite{Palais:1979rca}.
First applications in gravity can be found in Ref.~\cite{Deser:2003up}, where the consistency is ensured only if the components $g_{tt}$ and $g_{rr}$ of the metric are kept independent. Assuming in the ansatz the \textit{on-shell}
condition $g_{tt}g_{rr}=-1$\ can lead to inconsistencies \cite{Deser:2004yh} and the minisuperspace approximation does not apply in this case.

Let us focus on our solution \eqref{solution} and impose an ansatz on the fields that reduces the action, but reproduces the same dynamics.

%----------------------
\subparagraph{Ansatz.}
%---------------------

Consider the five-dimensional manifold $\mathcal{M}$ in spherical coordinates $x^{\mu }=(t,r,y^m)$, where $t$ is the time coordinate, $r$ is the radial one, and the coordinates $y^m$ parametrize the three-dimensional transversal manifold $\Sigma $ with unit-radius metric $\gamma
_{mn}(y)$ and the volume
\begin{equation}
\mathrm{Vol}(\Sigma )=\int \mathrm{d}^{3}y\,\sqrt{\gamma }\,,\qquad \gamma = \det [\gamma_{mn}].
\end{equation}
 In that way, the induced metric of the transversal section $t,r=const$ is $r^{2}\gamma_{mn}(y)$, and the asymptotic boundary of the spacetime, placed at $r=const\to \infty$, has the topology $\partial\mathcal{M} \simeq \mathbb{R} \times \Sigma$.

In AdS space, $\Sigma $ may have spherical, hyperbolic, or flat topology. In what follows, we focus on the spherical case, $\Sigma \simeq \mathbb{S}^{3}$.

Using the Lorentz indices decomposition $a=(0,1,i)$, the AdS part
of the gauge connection corresponds to the static, spherically symmetric black hole ansatz of the form
\begin{equation}
\begin{array}{llll}
e^{0}=Nf\,\mathrm{d}t\,, & e^{1}=\dfrac{\mathrm{d}r}{f}\,, & e^{i}=r\,\tilde{e}^{i}\,,\medskip  & \omega ^{0i}=0\,, \\ 
\omega ^{01}=\sigma \,\mathrm{d}t\,,\qquad  & \omega ^{1i}=-f\,\tilde{e}^{i},\qquad  & \omega ^{ij}=\epsilon ^{ijk}\omega _{k}\,,\qquad  & \omega
^{i}=\tilde{\omega}^{i}-\varphi \,\tilde{e}^{i}\,,
\end{array}
\label{e,w}
\end{equation}
where $N(r)$, $f(r)$, $\sigma (r)$, and $\varphi (r)$ are functions of the
radial coordinate only, and the vielbein and the spin connection on the sphere are $\tilde{e}^{i}=\tilde{e}_{m}^{i}(y)\mathrm{d}y^{m}$ and $\tilde{\omega}^{i}=\tilde{\omega}_{m}^{i}(y)\mathrm{d}y^{m}$, respectively. Note that $\gamma_{mn}=\delta_{ij}\, \tilde{e}^i_m \tilde{e}^j_n$, and our form of the spin connection $\omega^{ij}$ in \eqref{e,w} is already decomposed into the torsionless part $\tilde{\omega}^{ij}(e)$ plus the contorsion. 
This ansatz is good if it admits the solution \eqref{solution}, that is, $\hat f=f$, $\hat N=N$, $\hat \sigma =\sigma \equiv f( Nf)'$, and has the same conserved charges as in Sec.~\ref{review}, namely $\hat Q=Q$, $\hat E=E$.

Non-vanishing curvature components that follow from \eqref{e,w} are 
\begin{eqnarray}
    R^{01} &=&-\sigma'\,\mathrm{d}t\wedge \mathrm{d}r\,,\qquad R^{0i}=-f\sigma\,\mathrm{d}t\wedge \tilde{e}^{i}\,,  \notag \\
    R^{1i} &=&-f'\,\mathrm{d}r\wedge \tilde{e}^{i}-f\varphi \,\epsilon ^{ijk}\,\tilde{e}_{j}\wedge \tilde{e}_{k}\,,  \notag \\
    R^{ij} &=&\left( 1-\varphi ^{2}-f^{2}\right) \,\tilde{e}^{i}\wedge \tilde{e}^{j}-\varphi' \epsilon^{ijk}\,\mathrm{d}r\wedge \tilde{e}_{k}\,,
\label{R-ans}
\end{eqnarray}
and non-vanishing torsion components are
\begin{equation}
    T^{0}=\left[ \frac{\sigma }{f}-(Nf)'\right] \,\mathrm{d}t\wedge\mathrm{d}r\,,\qquad T^{i}=r\varphi \,\epsilon _{\ jk}^{i}\,\tilde{e}^{j}\wedge \tilde{e}^{k}\,.
\label{T-ans}
\end{equation}
The intrinsic curvature and torsion on the 3-sphere satisfy
\begin{eqnarray}
    \tilde{R}^{ij} & = &\epsilon ^{ijk}\mathrm{d}\tilde{\omega}_{k}-\tilde{\omega}^{i}\wedge \tilde{\omega}^{j}=\tilde{e}^{i}\wedge \tilde{e}^{j}\,, \notag \\
    \tilde{T}^{i} &= &\mathrm{d}\tilde{e}^{i}-\epsilon ^{ijk}\tilde{e}_{j}\wedge \tilde{\omega}_{k}=0\,.
\end{eqnarray}
Furthermore, the spherically symmetric ansatz of the internal field has the
form
\begin{eqnarray}
    \mathrm{U}(1)  &:&\quad A=A_{t}(r)\,dt\,,  \notag \\
    \mathrm{SU}(2) &:&\quad \mathcal{A}=(\tilde{\omega}^{i}-\phi \,\tilde{e}^{i})\mathbf{T}_{i}\,,\label{ans1}
\end{eqnarray}
with $A_{t}(r)$ and $\phi (r)$ being functions of the radial coordinate, and $\mathbf{T}_{i}$ is the anti-Hermitian $\mathrm{SU}(2)$ generator. 
We assume that all the fermions vanish on the solution.

The corresponding $\mathrm{U}(1)$ and $\mathrm{SU}(2)$ field strengths are, respectively, 
\begin{eqnarray} F&=&\mathrm{d}A=A'_t\,\mathrm{d}r\wedge \mathrm{d}t\,,  \notag \\
    \mathcal{F}^{i} &=&\mathrm{d}\mathcal{A}^{i}-\frac{1}{2}\,\epsilon ^{ijk}\mathcal{A}_{j}\mathcal{A}_{k}=-\phi' \,\mathrm{d}r\wedge \tilde{e}^{i}+\frac{1-\phi ^{2}}{2}\,\epsilon ^{ijk}\,\tilde{e}_{j}\wedge \tilde{e}_{k}\,. \label{ans}
\end{eqnarray}

The above ansatz depends on the reduced set of fields  $ \{N, f, \sigma, \varphi, A_{t}, \phi \}$, and is motivated by the structure of the expected solution \eqref{solution}. For example, the vielbein has a typical black hole form, the gauge field $\mathcal{A}$  is assumed to take the form of a soliton-like configuration, while the spin connection includes only specific components of the contorsion, rather than the most general set allowed by the symmetry. It is still necessary to verify the consistency of this ansatz -- that is, to ensure that the action principle, when applied to it, yields the expected equations of motion.

%----------------------
\subparagraph{Minisuperspace reduced action.}
%----------------------

Plugging in the chosen ansatz in the CS AdS action, we obtain a minisuperspace action $\mathring{I}[\Phi]$ as a 
function of the reduced fields $ \{N, f, \sigma, \varphi, A_{t}, \phi \}$.  For convenience, we also define the normalized five-dimensional volume element of dimension (length)$^2$,
\begin{equation}
    \mathrm{d}\Omega ^{5}\equiv \epsilon _{ijk}\,\mathrm{d}t\mathrm{d}r\tilde{e}^{i}\tilde{e}^{j}\tilde{e}^{k}=6\sqrt{\gamma }\,\mathrm{d}t\mathrm{d}r\mathrm{d}^{3}y\,= -(6\sqrt{\gamma }\,\mathrm{d}t\mathrm{d}^{3}y) \wedge \mathrm{d}r\,.
    \label{V.inv}
\end{equation}

The gravitational action in \eqref{action}--\eqref{L} is a polynomial in the Riemann curvature 2-form, such that each term in the polynomial can be evaluated as
\begin{eqnarray}
    \epsilon _{abcde}R^{ab}R^{cd}e^{e} &=&4\left[ \left( -\sigma -r\sigma'-Nf'\right) \left( 1-\varphi ^{2}-f^{2}\right) +2r\omega ff'+2Nf^{2}\varphi \varphi'\right] \,\mathrm{d}\Omega^{5}\,,  \notag \\
    \epsilon _{abcde}R^{ab}e^{c}e^{d}e^{e} &=&2\left[ -r^{3}\sigma'-3r^{2}\sigma -3r^{2}Nff'+3rN\left( 1-\varphi ^{2}-f^{2}\right) \right] \,\mathrm{d}\Omega ^{5}\,,  \notag \\
\epsilon _{abcde}e^{a}e^{b}e^{c}e^{d}e^{e} &=&20r^{3}N\,\mathrm{d}\Omega^{5}\,.
\end{eqnarray}
It yields to the pure gravitational term in the minisuperspace approximation,
\begin{equation}
    L_{\mathrm{G}}=\frac{1}{2\ell }\,\left[ \left( -\sigma -r\sigma' -Nff'+\frac{rN}{\ell ^{2}}\right) \Xi +2r\sigma ff'+2Nf^{2}\varphi \varphi'+\frac{2r^{3}\sigma'}{3\ell ^{2}}\right] \,\mathrm{d}\Omega ^{5}\,,
\end{equation}
where we defined the function 
\begin{equation}
    \Xi (r) = 1+\frac{r^{2}}{\ell ^{2}}-f^{2}-\varphi ^{2}\,. \label{Xi}
\end{equation}
whose on-shell value will give the charge $\Xi_0$ in \eqref{Xi-charge}. 

As regards the non-Abelian internal symmetry contribution to the action, 
as already discussed, we have  $L_{\mathrm{SU}(2)}=0$. This is a consequence of the fact that the CS action is defined as the Chern characteristic expressed in terms of the rank-3 symmetric invariant tensor, which is zero for the group $\mathrm{SU}(2)$, namely, $\mathrm{Tr}\left( \{ \mathbf{T}_{i},\mathbf{T}_{j}\}
\mathbf{T}_{k}\right) =d_{ijk}=0$.  Furthermore, given the expression of the $\mathrm{SU}(2)$ gauge field in the ansatz \eqref{ans1}--\eqref{ans}, with time component $\mathcal{A}_t=0$, it is not possible to construct a 5-form Lagrangian using only $\mathcal A$ and $\mathcal F$, which is a second reason for the vanishing of the non-Abelian action.

Finally, to compute the interaction term, we evaluate
\begin{eqnarray}
    T^{a}T_{a} &=&0\,,  \notag \\
    R^{ab}R_{ab} &=&\left[ 4ff'\varphi -2\left( 1-\varphi^{2}-f^{2}\right) \varphi'\right] \,\epsilon _{ijk}\,\mathrm{d}r \tilde{e}^{i}\tilde{e}^{j}\tilde{e}^{k}\,,  \notag \\
    R^{ab}e_{a}e_{b} &=&\left( -2r\varphi -r^{2}\varphi'\right) \,\epsilon _{ijk}\,\mathrm{d}r\tilde{e}^{i}\tilde{e}^{j}\tilde{e}^{k}\,,\notag \\
\mathcal{F}^{i}\mathcal{F}_{i} &=&-\left( 1-\phi ^{2}\right) \phi'\,\epsilon _{ijk}\,\mathrm{d}r\tilde{e}^{i}\tilde{e}^{j}\tilde{e}^{k}\,,
\end{eqnarray}
and obtain
%\gr{\begin{equation}    L_{\mathrm{int}}=-\frac{1}{4}\,\frac{\mathrm{d}}{\mathrm{d}r}\left( \Xi \varphi +\frac{2}{3}\,\varphi ^{3}-\phi +\frac{1}{3}\,\phi ^{3}\right) A_{t}\,\mathrm{d}\Omega^{5}  \,. %\label{int0} 
%\end{equation}}
\begin{equation}
    L_{\mathrm{int}}=-\frac{1}{4}\left( \Xi \varphi +\frac{2}{3}\,\varphi ^{3}-\phi +\frac{1}{3}\,\phi ^{3}\right) ' A_{t}\,\mathrm{d}\Omega^{5}  \,, \label{int0} 
\end{equation}
where the prime denotes the radial derivative.
Summing up the different contributions, the action acquires the form
\begin{equation}
   \mathring{I} = k\int \mathrm{d}\Omega ^{5}\mathring{L}(r)=-6k\int \mathrm{d}t\int \sqrt{\gamma }\,\mathrm{d}^{3}y\,\int \mathrm{d}r\mathring{L}(r)=-\Omega _{4}\int \mathrm{d}r\mathring{L}(r)\,,
\end{equation}
where $\Omega _{4} = 6k\Delta t\,\mathrm{Vol}(\Sigma )$, and the minus sign  takes into account that there is an additional minus sign in the definition of the normalized volume element $\mathrm{d}\Omega^5=-\mathrm{d}\Omega^4 \wedge \mathrm{d}r$ given by \eqref{V.inv}. Therefore, the minisuperspace action becomes
\begin{eqnarray}
    \mathring{I} &=&-\Omega _{4}\int \mathrm{d}r\left[ \frac{1}{2\ell }\,\left( -\sigma -r\sigma'-Nff'+\frac{rN}{\ell ^{2}}\right) \Xi +\frac{r^{3}\sigma'}{3\ell ^{3}}\right.   \notag \\
    &&\left. +\frac{r\sigma ff'}{\ell }+\frac{Nf^{2}\varphi \varphi'}{\ell }-\frac{1}{4}\,\left( \Xi \varphi +\frac{2}{3}\,\varphi^{3}-\phi +\frac{1}{3}\,\phi ^{3}\right)' A_{t}\right] .
\label{Imini}
\end{eqnarray}
In natural units, the fields $N$, $f$, $\varphi $, $\phi $,  are dimensionless, while $\sigma $ and $A_{t}$ have
dimensions of (length)$^{-1}$. In addition, considering that $\Xi $ is dimensionless, the expression in the square brackets has the units of (length)$^{-2}$. Finally, since $\Omega _{4}$ has the dimensions of a length, the action is dimensionless. This analysis uses the fact that the CS level $k$ is dimensionless.

%----------------------
\subparagraph{Equations of motion.}
%----------------------

Neglecting the boundary terms for now, the equations of motion %\gr{$\frac{\delta \mathring{I}}{\delta \Phi}=-\mathcal{E}_{\Phi}=0$}
$\delta \mathring{I} / \delta \Phi^\alpha =
-\Omega_4\,\mathcal{E}_{\alpha}=0$  have the following form:
\begin{eqnarray}
    \mathcal{E}_{N} &=&\frac{1}{2\ell}\left[\left( -ff'+\frac{r}{\ell ^{2}}\right) \Xi+2f^{2}\varphi \varphi']\right]\,, \notag \\
    \mathcal{E}_{f} &=&\frac 1{2\ell} \,f\left(N'\Xi +2N\varphi \varphi' -\ell A'_t \varphi\right)\,, \notag\\
    \mathcal{E}_{\sigma } &=&-\frac r{\ell}\,\varphi \varphi'\,,  \notag \\
    \mathcal{E}_{A_{t}} &=&-\frac 1{4}\left[2\varphi \left( \frac{r}{\ell ^{2}}-ff'\right) +\varphi'\Xi -\phi'\left( 1-\phi ^{2}\right) \right]\,,\notag \\
    \mathcal{E}_{\phi } &=&-\frac 1{4} \, A'_{t}\,\left( 1-\phi ^{2}\right) \,,  \notag \\
    \mathcal{E}_{\varphi } &=&\frac 1{2\ell}\left[-2\varphi \left( -\sigma -r\sigma' +\frac{rN}{\ell ^{2}}+(Nf)' f\right) +\frac{\ell }{2}\,\Xi A'_{t}\right]\,.
\end{eqnarray}
There are 6 equations and 6 unknown functions. They are nonlinear in $\Phi^\alpha$, thus they contain different branches. While $\mathcal{E}_{\sigma }=0$ can be directly solved as $\varphi
(r)=C=const$, the equation $\mathcal{E}_{\phi }=0$ distinguishes the cases $A'_t=0$
and $\phi ^{2}=1$. We are interested in the soliton solution where $\phi $ is not restricted, thus we will choose the first branch whose general
solution is $A_{t}(r)=A_{0}=const$. Then, the equation for $\mathcal{E}_{f}$ reduces to $\mathcal{E}_{f}=\frac{1}{2\ell }\,fN'\Xi =0$, which  is solved by  $N(r)=N_{0}=const$, since $f\neq 0$ and we are interested in the $\Xi $-charged solution with $\Xi \neq 0$. 
Using the above results, the equation $\mathcal{E}_{N}=0$  is solved by $f^{2}(r)=\frac{r^{2}}{\ell ^{2}}-e^2$, where $e^2 =const$. With this at
hand, the last two equations are straightforward to resolve, $\mathcal{E}
_{A_{t}}=0$ leads to $\phi (r)=B=const$, whereas $\mathcal{E}_{\varphi }=0$
gives $\sigma (r)=\frac{{N_{0}}\,r}{\ell ^{2}}+\dfrac{a}{r}$, with $a=const$.

To summarize, the general solution in the chosen branch reads
\begin{equation}
\begin{array}{lll}
    N=N_{0 }\,,\qquad  & \varphi =C\,,\qquad  & \sigma =\dfrac{{N_0}\,r}{\ell ^{2}}+\dfrac{a}{r}\,,\medskip  \\
    \phi =B\,, & A_{t}=A_{0 }\,, & f^{2}=\dfrac{r^{2}}{\ell ^{2}}-e^{2}\,,
\end{array}
\label{bh2}
\end{equation}
and the function $\Xi (r)$ on-shell is constant \eqref{Xi-charge},
\begin{equation}
    \Xi (r) = \Xi_0=e^{2}+1-C^{2}\,. 
\end{equation}
The integration constants correspond to the lapse ($N_{0 }$), axial
torsion ($C$), trace torsion ($a$), soliton amplitude ($B$), scalar
potential ($A_{0 }$) and extremality parameter ($e^{2}$), respectively, where $e^2$ can also be negative when there are no black holes (naked singularities and the global AdS). Note that all integration constants are dimensionless, except $A_0$, which has the units  (length)$^{-1}$.

Regarding the asymptotic conditions,  we find that, at  $r\to \infty$, the spacetime tends to global AdS geometry deformed by an axial torsion, so that the metric function and the trace torsion component behave as:
\begin{equation}
      N\to 1\,,\qquad f^2\to 1+\frac{r^{2}}{\ell ^{2}}-\mu\,,\qquad T^{0}_{tr}\to 0\,,\qquad \text{when }r\rightarrow\infty \,,\label{asympt}
\end{equation}
where the mass parameter $\mu$ measures the deviation from the global AdS space.\footnote{\label{DCBH}
The behavior $f^2_{\mathrm{CS}}\sim 1+\frac{r^{2}}{\ell ^{2}}-\mu$ of the metric functions is typical for Dimensionally Continued Black Holes that exist in any odd dimension \cite{Banados:1993ur}, and it differs drastically from Schwarzschild-like AdS$_D$ black holes behaving asymptotically as $f^2_{\mathrm{EH}}\sim 1+\frac{r^{2}}{\ell ^{2}}-\frac{\mu}{r^{D-3}}$.}
Since the torsion component  $T^{0}_{tr}=\frac{a}{rf}$ vanishes asymptotically, the behavior \eqref{asympt} does not put any restriction on the value of $a$. 
As regards the gauge field, the pure $U(1)$-gauge solution in the spherically symmetric ansatz implies the potential $A_t$ to be constant, so that in particular 
\begin{equation}
A_t \to const\,, \qquad  
\text{when }r\to \infty\,, \label{bc}
\end{equation} 
 while the other fields remain unrestricted. Imposing these boundary conditions on-shell implies
\begin{equation}
    N_{0 }=1\,,\qquad {\mu=e^2+1}\,,  \label{consts}
\end{equation}
while all other constants remain arbitrary. 

Comparing the obtained minisuperspace solution  \eqref{bh2} with the original $\Xi_0$-charged solution \eqref{solution} of CS AdS$_5$ supergravity without fermions, namely
\begin{equation}
\begin{array}{lll}
    \hat N=1\,,\qquad  & \hat \varphi =C\,,  & \hat \sigma =\dfrac{r}{\ell ^{2}}\,,\medskip  \\
    \hat \phi =B\,, & \hat A_{t}=const\,,\qquad & \hat f^{2}=\dfrac{r^{2}}{\ell ^{2}}-e^{2}\,,
\end{array}
\label{bh1}
\end{equation}
we conclude that \eqref{bh2} and \eqref{bh1} coincide, up to the trace torsion $a$ that was not considered in the original solution. The Abelian pure gauge field $\mathrm{d}\lambda$ becomes a constant in the spherically symmetric ansatz.
This means that the minisuperspace approximation of the CS AdS action yields a consistent reduction of the fields, provided suitable boundary conditions are imposed. It also enlarges the parameter space through the introduction of a new integration constant, $a$, associated with a solution with non-vanishing trace-component of the torsion.

In what follows, we allow for a constant $a$ to be non-vanishing. This extends the parameter space in such a way that all the charges leaving invariant the Lorentzian action can be constructed, together with the corresponding thermodynamic quantities in the Euclidean version of it. The original parameter space is recovered by setting $a=0$.

In addition, it can be seen from \eqref{bh2} that a natural field that incorporates the new solution with the integration constant $a$ is given on-shell by 
\begin{equation}
\psi(r)=\frac{a}{r}\,.    
\end{equation}
Thus, choosing this field as the fundamental one instead of $\sigma$, its off-shell expression has the form\footnote{The definition below has an on-shell equivalent form, $ \sigma(r)=\frac{r}{\ell ^{2}}\,N(r)+\psi(r)$. A direct computation shows that this
simplified relation reproduces the same results in all cases, including the
analysis of black hole thermodynamics. Nevertheless, since this equivalence
is not completely obvious, in particular because $\delta r=0$ when considering the
field variations, we shall work with the off-shell expression.}  
\begin{equation}
\sigma =f(fN)'+\psi \,,  \label{psi}
\end{equation}
where the first term comes from the torsionless component of the spin
connection $\hat \sigma$ in \eqref{solution}, namely, $\omega ^{01}=f(fN)'\diff t=\hat{\sigma}\,\diff t$. From
now on, we will treat $\psi(r)$ as an independent field instead of $\sigma(r)$, with the full set of the independent restricted fields being
\begin{equation}
\Phi ^{\alpha }=\{N,f,\psi ,\varphi ,A_{t},\phi \}\,.
\end{equation}
In that way, we have $\sigma = \hat \sigma + \psi$ and in the limit $\psi \to 0$ one recovers the original solution \eqref{bh1} with $\hat\sigma=\sigma$.  Physically, $\psi $ describes a non-trivial trace-component of the
torsion, $T^{0}_{tr}=\psi / f$.

%%%%%%%%%%%%%%%%%%%%%%%%%%%%%%%%%%%%%%%%%%
\section{ Conserved charges in minisuperspace}\label{lorentz}
%%%%%%%%%%%%%%%%%%%%%%%%%%%%%%%%%%%%%%%%%%

Let us compute the conserved charges for the solution \eqref{bh2}. As usual in the Noether theorem and Hamiltonian formalism, the information about conserved quantities is encoded in the boundary terms.
To see how they arise in our context, let us define a general static, spherically symmetric action in the radial foliation, with the Lagrangian that depends on some generalized coordinates $q_a(r)$ and  their `velocities' $q_a'=\frac{\mathrm{d} q_a}{\mathrm{d} r}$, that is, 
\begin{equation}
    I_\mathrm{reg}[q]=\Delta t\,L(q,q')\,,
     \label{fo}
\end{equation}
where the factor $\Delta t=t_{\mathrm{fin}}-t_{\mathrm{in}}$ is due to integration in time. The Lagrangian has the form
\begin{equation}
L(q,q')=\Omega_4 \int \mathrm{d}r\,\mathcal{L}(q(r),q'(r))  \,,  
\end{equation}
where the fields depend only on the radial coordinate. 
We require the action to be finite in the large-distance sector, which means that it has already been regularized by the addition of suitable counterterms to remove all the contributions diverging asymptotically.  
Varying in $q_a$, we get 
\begin{equation}
   \delta I_{\mathrm{reg}}=\Delta t\int \mathrm{d}r\,\left[ \Omega_3\,\mathcal{E}_a\, \delta q_a+\frac{\mathrm{d}}{\mathrm{d}r}\left( \frac{\partial \mathcal{L}}{\partial q'_a}\,\delta q_a\right) \right] .
\end{equation}
On-shell, that is, after imposing the Euler-Lagrange equations $\mathcal{E}_a=\frac{\partial \mathcal{L}}{\partial q_a}-\frac{\mathrm{d}}{\mathrm{d}r}\frac{\partial \mathcal{L}}{\partial q'_a}=0$, the action variation reduces to a  boundary term 
\begin{equation}
     \delta I^{{\mathrm{on-shell}}}_\mathrm{reg}=\Delta t\int \mathrm{d}r\,\frac{\mathrm{d}}{\mathrm{d}r}\left( \frac{\partial \mathcal{L}}{\partial q'_a}\,\delta q_a\right) = \Delta t \left[ \frac{\partial \mathcal{L}}{\partial q'_a}\,\delta q_a\right]_{\partial\mathcal{M}} =\Delta t\lim_{r\rightarrow \infty }\left( \frac{\partial \mathcal{L}}{\partial q'_a}\,\delta q_a\right).  \label{on-shell}
\end{equation}
Note that, in the above integral, where we applied Stokes' theorem, the boundary is defined at $r\to \infty$.  There is no boundary in $r=0$,  provided that the integrand is not singular at that point. 
 
The least action principle requires that the action variation \eqref{on-shell} vanishes when the  fields $q_a$ that satisfy $\frac{\partial \mathcal{L}}{\partial q'_a}\neq 0$ are kept fixed on the boundary $r\to \infty$, namely, $\delta q|_{\partial \mathcal{M}}=0$ (Dirichlet boundary condition).

To identify the pairs of canonically conjugate variables in the radial foliation, we notice that Hamiltonian actions, such as the CS one, have the general form
\begin{equation}
    I_\mathrm{reg}=\Delta t \int \mathrm{d}r\,\left( p_a q'_a-h\right) \,,
\end{equation}
with canonical momenta $p_a=\frac{\partial \mathcal{L}}{\partial q'_a}$ and a Hamiltonian density $h(p,q)$. In this particular case, the field equations are 
\begin{equation}
  \mathcal{E}_a=\frac{\partial p_b}{\partial q_a}\,q'_b-\frac{\partial h}{\partial q_a}-p'_a  \,. \label{EL}
\end{equation}
 Furthermore, applying \eqref{on-shell}, we get
\begin{equation}
    \delta I^{\mathrm{on-shell}}_\mathrm{reg}=\Delta t {\lim_{r\rightarrow \infty }}\,p_a\delta q_a\,,  \label{conjugate}
\end{equation}
from where the pairs of canonically conjugate variables $(p,q)$ are directly readable from the on-shell variation of the action with  Dirichlet boundary conditions.

On the other hand, we always have the freedom to add a finite boundary term $I_{\mathrm{fin}}$ to the action, amounting to a canonical transformation on a subset, say $\{ q_{\bar{a}} \}$, of the dynamical variables $\{ q_a \}$.  In particular, if we split our set of dynamical variables and momenta as $q_a=(q_{\bar{a}}, q_{\underline{a}}) $, we can modify the regularized action into 
\begin{equation}
I=I_{\mathrm{reg}}+I_{\mathrm{fin}}=I_{\mathrm{reg}}-\Delta t\int \mathrm{d}r\,\frac{\mathrm{d}}{\mathrm{d}r}\left( p_{\bar{a}} q_{\bar{a}}\right) \,, \label{conjugate1}
\end{equation}
such that the on-shell variation of the action \eqref{on-shell} changes to
\begin{equation}
    \delta I^{\mathrm{on-shell}}=\Delta t\lim_{r\rightarrow\infty }\,\left(p_{\underline{a}}\delta q_{\underline{a}}-q_{\bar{a}}\delta p_{\bar a}\right)\,. \label{conjugate2}
\end{equation}
This action satisfies the least action principle corresponding to the
momenta $p_{\bar a}$ kept fixed on the boundary, $\left.\delta p_{\bar a}\right|_{\partial \mathcal{M}}=0$ (Neumann boundary condition), while the unchanged coordinates $q_{\underline{a}}$ still satisfy the Dirichlet boundary condition, as in the original action \eqref{on-shell}.

It is worth mentioning that, since the on-shell fields $q_a$  in the Euclidean continuation describe the
chemical potentials, and the on-shell momenta $p_a$ describe the conserved charges, using Dirichlet vs.~Neumann boundary conditions corresponds to chemical potential vs.~charges fixed on the boundary, that is, it describes a change of the thermodynamic ensemble.

We will apply this procedure to the minisuperspace action to identify the pairs of conjugate variables among the fields $ \Phi^\alpha=\{N,f,\psi,\varphi,A_{t},\phi \} $.

%%%%%%%%%%%%%%%%%%%%%%%%%%%%%%%%%%%%%%%%%%
\subsection{On-shell variation of the action: counterterms and charges}
\label{Sec:Charges}
%%%%%%%%%%%%%%%%%%%%%%%%%%%%%%%%%%%%%%%%%%
From now on,  for the sake of simplicity, we will omit writing ``on-shell'' over the action.

The regularized minisuperspace action,
\begin{equation}
    I_\mathrm{reg}=\mathring{I}+I_\mathrm{ct}\,,
\label{Ireg0}
\end{equation}
includes, apart from the reduced bulk action $\mathring{I}$ given by \eqref{Imini},  also a surface counterterm  $I_\mathrm{ct}$ defined on the boundary $\partial \mathcal{M}$ located at $r=const\rightarrow \infty $.  
The counterterms remove long-distance ($r\to\infty$) divergences from the bulk action.

To determine the actual counterterms, we have to take into account all the boundary terms originally neglected when computing the equations of motion. In the chosen branch, these equations are equivalent to $N'=0$, $ff'=\frac{r}{\ell ^{2}}$, $\varphi'=0$, $\phi'=0$, $(r\psi)'=0$ and $A'_t=0$. Thus, starting from the action \eqref{Imini}, we get
\begin{eqnarray}
\delta \mathring{I} &=&-\Omega _{4}\int \mathrm{d}r\left[ \frac{1}{2\ell }\,\left( -\left( r\delta \sigma \right)'-\delta Nff'-N(f\delta f)'+\frac{r\delta N}{\ell ^{2}}\right) \Xi \right. \notag \\
&&+\frac{1}{\ell }\,\left( (r\sigma )'+Nff'-\frac{rN}{\ell^{2}}\right) \left( \rule[2pt]{0pt}{12pt}f\delta f+\varphi \delta \varphi
\right) +\frac{r^{3}\delta \sigma'}{3\ell ^{3}} \notag\\
&&+\frac{r\delta \sigma ff'}{\ell }+\frac{r\sigma (f\delta
f)'}{\ell }+\frac{\delta Nf^{2}\varphi \varphi'}{\ell }+\frac{N\delta f^{2}\varphi \varphi'}{\ell }+\frac{Nf^{2}\left( \varphi \delta \varphi \right)'}{\ell } \\
&&\left. -\frac{1}{4}\,\delta \left( \Xi \varphi +\frac{2}{3}\,\varphi^{3}-\phi +\frac{1}{3}\,\phi ^{3}\right)' A_{t}-\frac{1}{4}\,\left(
\Xi \varphi +\frac{2}{3}\,\varphi ^{3}-\phi +\frac{1}{3}\,\phi ^{3}\right)'\delta A_{t}\right]. \notag
\end{eqnarray}
The above expression depends on the variations of the fields,  $\delta \Phi^\alpha$, and their derivatives,  $\delta \Phi^{\prime \alpha}$. However, according to the action principle, the equations of motion, $-\mathcal{E}_\alpha$, appear as the coefficients of $\delta \Phi^\alpha$. 
Therefore, to isolate them properly, any derivatives acting on the field variations must first be eliminated by integration by parts. As a result, the variation of the action on-shell becomes a total derivative,
\begin{eqnarray}
\delta \mathring{I} &=&-\Omega _{4}\int \mathrm{d}r\,\frac{\mathrm{d}}{%
\mathrm{d}r}\,\left[ \left( \frac{r^{3}}{3\ell ^{3}}-\frac{r\Xi }{2\ell }%
\right) \,\delta \sigma +\left( r\sigma -\frac{N\Xi }{2}\right) \frac{%
\,f\delta f}{\ell }\right.   \notag \\
&&+\left. \frac{Nf^{2}}{\ell }\,\varphi \delta \varphi -\frac{1}{4}\,\delta
\left( \Xi \varphi +\frac{2}{3}\,\varphi ^{3}-\phi +\frac{1}{3}\,\phi
^{3}\right) A_{t}\right] ,
\end{eqnarray}
up to additional terms which are proportional to the equations of motion, and thus vanish. For example, the coefficient $\frac{1}{\ell }\,\left[ (r\sigma )'-N\left( ff'+\frac{r}{\ell ^{2}}\right) \right]$, multiplying $\varphi \delta \varphi $, is zero on-shell because $ff'=\frac{r}{\ell ^{2}}$
and $(r\sigma )'=\frac{2rN}{\ell ^{2}}$, and similarly for other  coefficients.

Using  the definition \eqref{psi} to introduce the field $\psi$, we get
\begin{eqnarray}
\delta \mathring{I} &=&-\Omega _{4}\int \mathrm{d}r\,\frac{\mathrm{d}}{\mathrm{d}r}\, \left[ \left( \frac{r^{3}}{3\ell ^{3}}-\frac{r\Xi }{2\ell }\right)  \left( \rule[2pt]{0pt}{11pt} \delta(f(fN)')+ \delta\psi \right) +\left( r\psi
+rf(fN)'-\frac{N\Xi }{2}\right) \frac{\,f\delta f}{\ell }\right.  
\notag \\
&&+\left. \frac{Nf^{2}}{\ell }\,\varphi \delta \varphi -\frac{1}{4}\,\delta
\left( \Xi \varphi +\frac{2}{3}\,\varphi ^{3}-\phi +\frac{1}{3}\,\phi
^{3}\right) A_{t}\right] .
\label{deltaI}
\end{eqnarray}
After solving the integral in \eqref{deltaI}, the on-shell variation takes the form
\begin{eqnarray}
\delta \mathring{I} &=&-\Omega _{4}\lim_{r\rightarrow \infty }\left[ \left( 
\frac{r^{3}}{3\ell ^{3}}-\frac{r\Xi }{2\ell }\right) \left( 
\rule[2pt]{0pt}{11pt}\delta (f(fN)')+\delta \psi \right) +\left(
r\psi +rf(fN)'-\frac{N\Xi }{2}\right) \frac{f\delta f}{\ell }\right.   \notag \\
&&+\left. \frac{Nf^{2}}{\ell }\,\varphi \delta \varphi -\frac{1}{4}\,\delta
\left( \Xi \varphi +\frac{2}{3}\,\varphi ^{3}-\phi +\frac{1}{3}\,\phi
^{3}\right) A_{t}\right] \,.
\end{eqnarray}
Plugging in the solution \eqref{bh2}  and varying only the  parameters (the integration constants $e$, $N_0$, $a$, $C$, $A_0$ and $B$),  while $\delta r=0$, we get 
\begin{eqnarray}
    \delta \mathring{I} &=&-\Omega _{4}\lim_{r\rightarrow \infty }\left[ \left(\frac{r^{2}}{3\ell ^{3}}-\frac{\Xi_0}{2\ell }\right) \delta a- \left( \dfrac{N_{0 }r^{2}}{\ell ^{2}}+a-\dfrac{N_{0 }\Xi _{0 }}{2}\right) \frac{\delta e^{2}}{2\ell }+\delta N_0\,\dfrac{r^2}{\ell^3}\left(\dfrac{r^2}{3\ell^2}-\frac{\Xi_0}{2} \right) \right.  \nonumber  \\
    &&+\left. \frac{N_{0 }}{\ell }\left( \dfrac{r^{2}}{\ell ^{2}}-e^{2}\right) C\delta C-\frac{1}{4}\,\delta \left( \Xi _{0 }C+\frac{2}{3}\,C^{3}-B+\frac{1}{3}\,B^{3}\right) A_{0 }\right]\,, \label{Ionew}
\end{eqnarray}
where
\begin{equation}
    \Xi _{0}=1+e^{2}-C^{2}\,,\qquad \delta \Xi_0= \delta e^2 -\delta C^2\,.
\end{equation}

%%%%%%%%%%%%%%%%%%%%%%%%%%%%%
\subparagraph{Counterterms.}
%%%%%%%%%%%%%%%%%%%%%%%%%%%%%

To remove the divergences, the action must be supplemented with additional boundary terms, namely counterterms, so that all divergent contributions can be written as total variations. For instance, the divergent term proportional to $\delta a$ is
\begin{equation}
    \frac{{r^2}}{3\ell ^{3}}\,\delta a =\delta \left( \frac{{r^2}a}{3\ell ^{3}}\right) \,,
\label{count-a}
\end{equation}
while along $\delta N_0$ we find
\begin{equation}
    \delta N_0\,\dfrac{r^2}{2\ell^3}\left(\dfrac{2\,r^2}{3\ell^2}- \Xi_0\right) -N_0\,\dfrac{r^2}{2\ell^3}\left(\delta e^2 -\delta C^2\right)=\delta\left[N_0\,\dfrac{r^2}{2\ell^3}\left(\dfrac{2\,r^2}{3\ell^2}- \Xi_0\right)\right].\label{count}
\end{equation}
After the addition of the above boundary terms, the variation of the obtained regularized action only includes integrable expressions, i.e., written as total variations. Following our notation introduced in \eqref{Ireg0}, the terms \eqref{count-a} and \eqref{count} define the counterterm $-\delta I_{\mathrm{ct}}$ which, integrated, becomes
\begin{equation}
    I_{\mathrm{ct}}=\Omega _{4}\lim_{r\rightarrow \infty }\left[ \frac{r^2a}{3\ell^3}+ \frac{r^2N_{0}}{2\ell^3}\left(\frac{2r^2}{3\ell^{2}}-\Xi_0\right)\right]\,.  \label{Ict} 
\end{equation}
In terms of the fundamental fields, this surface term reads
\begin{equation}
I_{\mathrm{ct}}=\Omega _{4}\int \mathrm{d}r\,\frac{\mathrm{d}}{\mathrm{d}r}\,\left[ \frac{r^{3}\psi }{3\ell ^{3}} + rf(fN)' \left( \frac{r^{2}}{3\ell ^{3}} -\frac{\Xi }{2\ell }\right) \right] \,, \label{Ictfundnew}
\end{equation}
and it makes  $I_\mathrm{reg}=\mathring{I}+I_\mathrm{ct}$  finite at the boundary. 

It is worth noticing that the same result can be obtained directly from 
the action \eqref{Imini}, without varying it. 
Indeed, on-shell, the action becomes a total derivative that, integrated, takes the form
\begin{equation}
\mathring{I}_{\mathrm{class}}=-\Omega _{4}\lim_{r\rightarrow \infty }\left[ \frac{r}{\ell }%
\left( -\frac{\Xi }{2}+\frac{r^{2}}{3\ell ^{2}}\right) \left( 
\rule[2pt]{0pt}{12pt}f(fN)'+\psi \right) -\frac{1}{4}\,\left( \Xi
\varphi +\frac{2}{3}\,\varphi ^{3}-\phi +\frac{1}{3}\,\phi ^{3}\right) A_{t}
\right] \,.
\end{equation} 
Its evaluation on the solution,
\begin{equation}
\mathring{I}_{\mathrm{class}}=-\Omega _{4}\left[ \lim_{r\rightarrow \infty }\left( -\frac{\Xi
_{0}}{2}+\frac{r^{2}}{3\ell ^{2}}\right) \left( \frac{r^{3}}{\ell ^{3}}+%
\frac{a}{\ell }\right) -\frac{1}{4}\,\left( \Xi _{0}C+\frac{2}{3}\,C^{3}-B+%
\frac{1}{3}\,B^{3}\right) A_{0}\right] \,,
\end{equation}%
shows that the classical action is divergent.  The divergent part is identified as
\begin{equation}
-I_{\mathrm{ct}}=-\Omega _{4}\lim_{r\rightarrow \infty }\left[ \left( -\frac{%
\Xi _{0}}{2}+\frac{r^{2}}{3\ell ^{2}}\right) \frac{r^{3}}{\ell ^{3}}+\frac{%
ar^{2}}{3\ell ^{3}}\right] \,,
\end{equation}%
or, expressed in terms of $\Phi ^{\alpha }$, as
\begin{equation}
I_{\mathrm{ct}}=\Omega _{4} \int \mathrm{d}r\,\frac{\mathrm{d}}{\mathrm{d}r}\,\left[ rf(fN)^{\prime
}\left( -\frac{\Xi }{2\ell }+\frac{r^{2}}{3\ell ^{3}}\right) +\frac{%
r^{3}\psi }{3\ell ^{3}}\right] \,,
\end{equation}
which coincides with the previous result \eqref{Ictfundnew}.

Regarding the finite part, it does not necessarily coincide with $I_\mathrm{reg}$, since it contains ambiguities. In particular, one may always add to the action some terms proportional to the equations of motion written as total derivatives. On the other hand, the variation of the action does not suffer from this type of ambiguity.
For this reason, $\delta \mathring{I}$ is more suitable for the analysis, as it relates the on-shell variation of the action to the boundary conditions imposed on the fields or parameters. Indeed, the addition of a finite counterterm should modify both the action and the boundary conditions of the fields.

%%%%%%%%%%%%%%%%%%%%%%%%%%%%%
\subparagraph{Charges.}
%%%%%%%%%%%%%%%%%%%%%%%%%%%%%

Returning to \eqref{Ionew} to obtain the regularized action \eqref{Ireg0}, we arrive at the result
\begin{eqnarray}
    \delta I_\mathrm{reg}=\delta(\mathring{I}+I_\mathrm{ct}) &=&\Omega _{4}\left[ \delta a\,\frac{\Xi_{0 }}{2\ell } +\frac{ \delta e^{2}}{4\ell }\,\left(2a- N_{0}\Xi _{0 }\right)+\delta C^2\,\frac{N_{0}}{2\ell }e^{2}\right.   \notag\\
    &&\hskip 5mm +\left.  \,\frac{1}{4}\,A_{0}\,\delta \left( \Xi _{0 }C+\frac{2}{3}\,C^{3}- B  +\frac{1}{3}\,B^{3}\right)\right] \,.\label{dL}
\end{eqnarray}
We have 6 integration constants $(N_0,e,B,C,a,A_0)$ associated with 6 fields $\Phi^\alpha$ that should describe 3  canonically conjugate pairs of variables. To identify them, we use the method described at the beginning of this subsection (see the discussion above eq.~\eqref{conjugate2}).  
It is known that $N$ and $E$ are conjugate variables\footnote{According to Ref.~\cite{Crisostomo:2000bb}, the Hamiltonian lapse function $N^{\perp}$ is defined in the ADM formalism as $N^{\perp}=(-g^{tt})^{- \frac 12}$, while the variable conjugate to the energy is $N=N^{\perp}/f$. This notation is consistent with our metric \eqref{metric}.} \cite{Crisostomo:2000bb}. 
On the other hand, the potential $A_{0}$ is conjugate to some $\mathrm{U}(1)$ charge $Q$,  and let us call $P$ the momentum conjugate to the trace torsion parameter $a$. Thus, following the prescription \eqref{conjugate2}, the variation of the regularized action should be cast into the form 
\begin{equation}
    \delta I_{\text{expected}}=  \Delta t\,(-\delta E\,N_{0 }-A_0\, \delta Q+ P \delta a)\,,
 \label{should.be}
\end{equation}
which divides the 6 variables $\Phi^\alpha$ into three generalized coordinates $q_a=(N_0,A_0,a)$ and three conjugate momenta $p_a=(E,Q,P)$. The coordinates come from the fields $N$, $A_t$ and $\sigma$, while the momenta correspond to conserved charges, being combinations of the integration constants $e^2$, $C$ and $B$. The action should then vanish upon imposing Dirichlet boundary conditions on the field $a$, and Neumann boundary conditions on the momenta $E$ and $Q$.

However, the expression \eqref{dL} that we found is not of the form \eqref{should.be}, because it contains, besides contributions in $\delta a$, also terms with $a$, which do not vanish upon imposing Dirichlet boundary conditions on $a$.  Thus, to put our regularized action in a proper form, we have to add a finite counterterm  $I_{\mathrm{fin}}^{(a)}$, which will change the boundary conditions. As we will see, in the Euclidean continuation of the spacetime, adding this finite term will correspond to changing the thermodynamical ensemble.

Thus, we define
\begin{equation}
    I^{(a)}_{\mathrm{fin}}=-\Delta t \, \Omega_3\, \frac{e^2 a}{2\ell}\,, \label{fin1}
\end{equation}
on the asymptotic boundary, written in terms of the fundamental fields in the bulk as
\begin{equation}
I_{\mathrm{fin}}^{(a)}=-\Omega _{4}\,\int \mathrm{d}r\,\frac{\mathrm{d}}{\mathrm{d}r}\left[ \frac{r\psi} {2\ell }\left( \frac{r^{2}}{\ell ^{2}}-f^{2}\right) \right] \,. \label{Ifin1}
\end{equation}

With this at hand, the total action reads
\begin{equation}
I^{(a)}=I_{\mathrm{reg}}+I^{(a)}_{\mathrm{fin}}=\mathring{I}+I_{\mathrm{ct}}+I^{(a)}_{\mathrm{fin}}\,.
\label{Ireg1}
\end{equation}
Its variation indeed has the form \eqref{should.be},
\begin{eqnarray}
    \delta I^{(a)} &=&\Delta t \, \Omega _{3}\left[ N_{0}\left( -\frac{\delta e^{2}}{4\ell }\,\Xi _{0}+\delta C^{2}\,\frac{e^{2}}{2\ell }\right) -\left( \frac{e^{2}}{2\ell }-\frac{\Xi _{0}}{2\ell }\right) \delta a\right.   \notag \\
    &&+\left. \frac{1}{4}\,A_{0}\,\delta \left( \Xi _{0}C+\frac{2}{3}\,C^{3}+B-\frac{1}{3}\,B^{3}\right) \right] \,, \label{dL(a)}
\end{eqnarray} 
 where we used the definition $\Omega_4=\Delta t\,\Omega_3$. Thus, we recognize the set of conjugate variables from \eqref{should.be}, that is
\begin{equation}
    \delta I^{(a)}=\Delta t\,(-\delta E\,N_{0}-A_{0}\,\delta Q+P\delta a)\,.
    \label{varIa}
\end{equation}

Comparing the last two formulas, \eqref{dL(a)} and \eqref{varIa}, we can identify the variation of the energy as the term proportional to $N_0$,
\begin{equation}
    \delta E=\frac{\Omega _3}{4\ell }\,\left(\Xi _{0 }\,\delta e^{2}-2e^{2}\delta C^2 \right) \,, \label{E}
\end{equation}
where $\Omega_3=12k\pi^2$.
This matches\footnote{With the following change of notation compared to \cite{Andrianopoli:2021qli}:  $6k\mathrm{Vol}(\Sigma) \to \Omega_3$ and $p^2 \to e^2$.} the variation of the Hamiltonian energy \eqref{hatE} obtained in \cite{Andrianopoli:2021qli}, namely, $\delta E=\delta \hat E$. 

Furthermore, focusing on the term proportional to the variation of $A_0$, we find the $U(1)$ charge 
\begin{eqnarray}
    Q 
    &=&-\frac{\Omega _3}{4}\left( e^{2}C+C-\frac 13\,C^{3}-B+\frac{1}{3}\,B^{3}\right) \,, 
    \label{Qmini}
\end{eqnarray}
which again matches the expression \eqref{hatQ} found in \cite{Andrianopoli:2021qli}, confirming that $Q=\hat Q$.

Finally, the momentum $P$ is identified from the term proportional to $\delta a$ as
\begin{equation}
    P=-\frac{\Omega_3}{2\ell}\, (C^2-1)\,.
    \label{Pmini}
\end{equation}

 To summarize,  we found the explicit expressions for all conjugate quantities, except for the energy, where we obtained only its variation, which  is not integrable in general because, from \eqref{E}, 
\begin{eqnarray}
    \delta E=\frac{\Omega_3}{4\ell}  \left[ \delta \left(  e^2\Xi_0-\frac{e^4}{2} \right) - e^2 \delta C^2 \right].
\end{eqnarray}
It coincides with $\delta\hat E$ given by \eqref{hatE}. For the study of black hole thermodynamics, which we will do in the next section, knowing $\delta E$ is sufficient. However, having the explicit expression for $E$  allows for a direct comparison with the results obtained by other methods.
In \cite{Andrianopoli:2021qli}, the energy was integrated by imposing  $\delta C=0$,  a condition that is well justified when $C$ is a quantized quantity. We assume that $a \neq 0$ does not modify the topological charges \eqref{n12} because they are computed on the surface $\Sigma$ at $\mathrm{d}t=0$, where $\omega^{01}|_{\Sigma} =0$ and $T^0|_{\Sigma}=0$. In this case,  \eqref{E} can be integrated out as
\begin{equation}
    E|_{\delta C=0}= \frac{\Omega _3}{4\ell }\,\left( \Xi _{0 }-\frac{e^2}{2}\right)e^2\,,\label{pairs}
\end{equation}
where we have chosen a $C$-dependent additive constant such that the energy vanishes for the extremal black hole $e^2=0$. 

In the next section, we treat $C$ as a thermodynamic variable ($\delta C \neq 0$) in order to study its thermodynamic behavior, and set $\delta C=0$ only when integrating out the charges or the entropy.

%%%%%%%%%%%%%%%%%%%%%%%%%%%%%%%%
\subsection{Vacuum energy}
\label{vacuum-energy-section}
%%%%%%%%%%%%%%%%%%%%%%%%%%%%%%%%

In odd dimensions, the global AdS space without torsion, characterized by $e^{2}=-1$ and $C=0$, has  vacuum energy obtained from \eqref{pairs} as
\begin{equation}
    E_{\mathrm{vac}}(0)=-\frac{\Omega _{3}}{8\ell }\,.
    \label{AdSvacuum}
\end{equation}
Compared with the result in General Relativity, one observes an overall sign difference, using the notation introduced in footnotes \ref{radius} and \ref{BH scan} and setting $k=\frac{\ell ^{3}}{8\pi G_{N}
}$. In five-dimensional asymptotically AdS Einstein gravity, the vacuum energy of global AdS is nonzero and positive: it is the mass assigned to pure AdS for boundary topology $\mathbb{S}^{3}\times \mathbb{R}$, as obtained from the holographic Balasubramanian--Kraus stress tensor \cite{Emparan:1999pm,Mora:2004rx}. In CS AdS gravity, instead, the vacuum energy is negative \cite{Banados:2004zt}. This sign reversal is natural, since CS AdS black holes arise as the higher-dimensional continuation of the three-dimensional BTZ black hole, in the sense that they retain the same form of the metric function, $f^{2}=\frac{r^{2}}{\ell ^{2}}-M$ (see footnote \ref{DCBH}). For the BTZ solution, global AdS$_{3}$ is obtained at $M=-1$, so the vacuum already has negative energy. Accordingly, the result \eqref{AdSvacuum} reproduces the standard vacuum energy of CS AdS$_{5}$ gravity.

In our analysis, when torsion is included, i.e., when   $C\neq 0$, the vacuum energy acquires an additional positive contribution due to the presence of the axial torsion that changes the vacuum geometry into a solitonic background (see \eqref{solution}). As a result, the total energy becomes
\begin{equation}
E_{\mathrm{{vac}}}(C)=E_{\mathrm{{vac}}}(0)+\frac{\Omega _{3}}{{4}\ell }
\,C^{2}=|E_{\mathrm{{vac}}}(0)|(-1+2C^{2})\,.
\label{Evac}
\end{equation}
Therefore, the axial torsion increases the negative vacuum energy of AdS space \eqref{AdSvacuum} by a term quadratic in the torsion parameter $C$.

%%%%%%%%%%%%%%%%%%%%%%%%%%%%%%%%
\paragraph{Casimir energy.}
%%%%%%%%%%%%%%%%%%%%%%%%%%%%%%%%%

In the framework of the AdS/CFT correspondence, the quantity $E|_{\delta C=0}$ given by \eqref{pairs} represents the energy of the conformal field theory (CFT) that is holographically dual to the bulk gravitational theory. The vacuum energy of global AdS space in five dimensions without torsion is identified with the four-dimensional Casimir energy of the CFT defined on the spacetime $\mathbb{R}\times \mathbb{S}^{3}$, that is, $E_{\mathrm{vac}}^{5d}=E_{\mathrm{Casimir}}^{4d}$. The expression \eqref{AdSvacuum} corresponds to the well-known Casimir energy of a CFT in a static curved background.

Yet another way to obtain the Casimir energy with torsion and test the result \eqref{Evac} is to use the holographic stress tensor in a CFT defined on $\Sigma \simeq \mathbb{S}^{3}$, as
discussed in \cite{Banados:2006fe}. It is computed as a 1-point function, 
\begin{equation}
E_{\mathrm{Casimir}}=\mathrm{Vol}(\Sigma )\,\left\langle \mathcal{T}_{tt}\right\rangle \,, 
\label{Casimir}
\end{equation}
of the holographic stress tensor $\left\langle \mathcal{T}_{\ \ A}^{\alpha}\right\rangle $ in CFT$_{4}$, 
expressed entirely in terms of the gravitational sources $e_{(0)}^{A}=e_{(0)\alpha}^{A}\,\diff x^{\alpha}$ and $\omega _{(0)}^{AB}=\omega
_{(0)\alpha}^{AB}\,\diff x^{\alpha}$ coming from the five-dimensional bulk. Here, $A$, $B$ are the boundary Lorentz indices and $\alpha,\beta$ are the corresponding 4D
spacetime indices. In our notation, the decomposition of the indices to lower dimensions is $a=(1,A)$ and $A=(0,i)$, which on the spacetime manifold corresponds, respectively, to $
x^{\mu }=(r,x^{\alpha})$ and $x^{\alpha}=(t,y^m)$.  
Note that energy and temporal component of the stress tensor have dimensions (length)$^{-1}$, $e^0_{(0)t}$ and $\omega _{(0)m}^{ij}$ are dimensionless (thus the curvature $R_{(0)mn}^{ij}$ as well), while $e_{(0)m}^{i}$ have dimensions of length.

The holographic stress tensor $\left\langle \mathcal{T}_{\alpha\beta}\right\rangle =e_{(0)\alpha}^{A}\,g_{(0)\beta \gamma}\left\langle \mathcal{T}_{\ \
A}^{\gamma}\right\rangle $ in CFT$_{4}$ is computed in CS AdS$_5$ gravity using the holographic techniques as \cite{Banados:2006fe}
\begin{equation}
\left\langle \mathcal{T}_{\ \ A}^{\alpha}\right\rangle =\frac{k}{\ell^2\sqrt{-g_{(0)}}}\,\epsilon ^{\alpha\beta\gamma\delta}\epsilon _{ABCD}\,\left( \frac{1}{2}\,R_{(0)\beta\gamma}^{BC}+\frac{2}{\ell ^{2}}\,e_{(0)\beta}^{B}k_{\    \gamma}^{C}\right) k_{\ \delta}^{D}\,,
\label{holT}
\end{equation}
where  $R_{(0)}^{AB}=\mathrm{d}\omega _{(0)}^{AB}+\omega _{(0)}^{AC} \omega _{(0)C}^{\ \  \ \ B}$ is the
intrinsic curvature 2-form of the boundary, and $k^{A}=k_{\ \alpha}^{A}\,\diff
x^{\alpha}$ is a 1-form that appears in the sub-leading order of the asymptotic
expansion of the fields. The components $k_m^{i}$ have dimensions of length. In the Fefferman-Graham-like frame $x^{\alpha}=(\rho
,y^m)$, where the new radial coordinate $\rho$ is dimensionless, with the boundary placed at $\rho =0$, this expansion reads \cite{Banados:2006fe}
\begin{equation}
\begin{array}[b]{llll}
e^{1} & =-\dfrac{\ell \diff\rho }{2\rho }\,,\medskip \qquad  & e^{A} & =\dfrac{1}{\sqrt{\rho }}\,e_{(0)}^{A}+\sqrt{\rho }\,k^{A}, \\
\omega ^{AB} & =\omega _{(0)}^{AB}\,, & \omega ^{A1} & =\dfrac{1}{\ell \sqrt{\rho }}\,e_{(0)}^{A}-\dfrac{\sqrt{\rho }}{\ell }\,k^{A}\,.
\end{array}
\end{equation}
To compute the holographic stress tensor, we have to change the coordinate frame of the black hole \eqref{metric} from the Schwarzschild-like  one $(t,r,y^m)$  to the Fefferman-Graham-like one $(t,\rho,y^m)$, by changing the radial coordinate as
\begin{equation}
     r=\frac{\ell}{\sqrt{\rho}}+\frac{\ell e^2}{4}\sqrt{\rho}\,. 
\end{equation}
Replacing $r(\rho )$ in the solution \eqref{e,w}, \eqref{bh2},  when $N_{0}=1$ and $a=0$ and in the asymptotic limit $\rho \rightarrow 0$, we can identify the boundary CFT quantities in terms of the black hole parameters: 
\begin{equation}
\begin{array}[b]{llll}
e_{(0)}^{0} & =\mathrm{d}t\,,\qquad \medskip  & k^{0} & =-\dfrac{e^{2}}{4}\,%
\mathrm{d}t\,, \\
e_{(0)}^{i} & =\ell \,\tilde{e}^{i}\,,\medskip  & k^{i} & =\dfrac{\ell e^{2}%
}{4}\,\tilde{e}^{i}\,, \\
\omega _{(0)}^{0i} & =0\,, & \omega _{(0)}^{ij} & =\epsilon ^{ijk}\left(
\tilde{\omega}_{k}-C\,\tilde{e}_{k}\right) \,.%
\end{array}
\label{sources}
\end{equation}
In consequence, we  find the boundary metric 
\begin{equation}
g_{(0)\alpha\beta}\,\mathrm{d}x^{m}\mathrm{d}x^{n}=-\mathrm{d}t^{2}+\ell ^{2}\,\gamma_{mn}\,\mathrm{d}y^m\mathrm{d}y^n\,,  \qquad \sqrt{-g_{(0)}}=\sqrt{\gamma}\,,
\end{equation} 
 and compute the boundary curvature
\begin{equation}
R_{(0)}^{ij}=\left( 1-C^{2}\right) \,\tilde{e}^{i}\tilde{e}^{j}\,.
\label{boundaryR}
\end{equation}
Then, the Casimir energy \eqref{Casimir} is obtained using \eqref{holT}, \eqref{sources} and \eqref{boundaryR}. The energy density has the form 
\begin{equation}
\left\langle \mathcal{T}_{tt}\right\rangle =-\left\langle \mathcal{T}_{\ \
0}^{t}\right\rangle =\frac{3ke^{2}}{2\ell}\,\left( \dfrac{e^{2}}{2}
+1-C^{2}\right) ,
\end{equation}
where we applied the identity
\begin{equation}
\frac{1}{\sqrt{\gamma}}\,\epsilon ^{tmns}\epsilon _{0ijk}\,\tilde{e}_{m}^{i}
\tilde{e}_{n}^{j}\tilde{e}_{s}^{k}=-\epsilon
^{ijk}\epsilon _{ijk}=-6\,.
\end{equation}
Comparing this result with \eqref{Evac}, we get that the vacuum energy of AdS space with torsion matches the Casimir energy in the holographic CFT$_4$,  namely, $E_{\mathrm{vac}}=E_{\mathrm{Casimir}}$.

%%%%%%%%%%%%%%%%%%%%%%%%%%%%%%%
\section{Entropy and the first law of thermodynamics}
\label{euclid}
%%%%%%%%%%%%%%%%%%%%%%%%%%%%%%%

In the previous sections, we used the minisuperspace action in a spacetime with Lorentzian signature to obtain information about the conserved charges of the theory.

In this section, we perform a Euclidean continuation of the spacetime in order to study the statistical properties of the theory and investigate the thermodynamic behavior of the black hole-soliton system. We focus on the non-extremal black holes with $e^2 > 0$, which possess a finite Hawking temperature, $T>0$.

For the static action $I=\Delta t\,L$ of the form \eqref{fo}, the Wick rotation $t\to \mathrm{i}\tau $ changes the Lorentzian partition function into the Euclidean one,
\begin{equation}
\mathrm{e}^{\mathrm{i}I}=\mathrm{e}^{\mathrm{i}\Delta t\,L}\rightarrow
\mathrm{e}^{-I^{E}}=\mathrm{e}^{\mathrm{i}(\mathrm{i}\Delta \tau )\,L}=\mathrm{e}^{-\beta G}\,,
\label{partition} 
\end{equation}
where the Euclidean period is identified with the inverse temperature, $\Delta \tau
=\beta =1/T$, when the Boltzmann constant is set to 1.
Then the Euclidean Lagrangian on-shell becomes a free energy, $G=L={\Omega_3}\int
\mathrm{d}r\,\mathcal{L}(r)$. It implies that the Euclidean action on-shell is also related to the free energy,
\begin{equation}
G=TI^{E}\,.
\label{G=TI}
\end{equation}
This relation holds provided the action is finite and satisfies a suitable variational principle.

The first law of thermodynamics reads 
\begin{equation}
\delta E=T\delta S+\sum\limits_{a} Q_a \delta \mu_a\,,\qquad
E=E(S,\mu_a)\,,
\label{1law}
\end{equation}
where $S$ is the entropy of the system, $\mu_a$ are intensive variables (chemical potentials, identified with the generalized coordinates $q_a$) and $Q_a$ are extensive variables (charges, corresponding to generalized canonical momenta $p_a$).

We can change the microcanonical ensemble, with thermodynamic potential $E(S,\mu_a)$, to a different one, by performing a Legendre transformation with respect to a subset of the charges, denoted $Q_{\bar a}$, and the entropy, while leaving the remaining charges $Q_{\underline{a}}$ untransformed. This defines a decomposition of the charge indices as $a=(\bar a,\underline{a})$.\footnote{The notation is the same as that introduced in eqs.~\eqref{conjugate1} and \eqref{conjugate2} for the Lorentzian case.} The resulting thermodynamic potential
\begin{equation}
G = E - TS - \sum\limits_{\bar{a}}\mu _{\bar{a}}Q_{\bar{a}}\,,\qquad
G=G(T,\mu _{\underline{a}},Q_{\bar{a}})\,,
\label{G}
\end{equation}
defines a (grand) canonical ensemble,
with the corresponding first law of thermodynamics given by
\begin{equation}
\delta G=-S\,\delta T+\sum\limits_{\underline{a}}Q_{\underline{a}
}\delta \mu _{\underline{a}}-\sum\limits_{\bar{a}}\mu _{\bar{a}}\delta Q_{\bar{a}}\,.  
\label{1st law}
\end{equation}

%%%%%%%%%%%%%%%%%%%%%%%%%%%%%%%%%%%%%%%%%%%%%%%%
\subparagraph{Boundary terms and ensembles.}
%%%%%%%%%%%%%%%%%%%%%%%%%%%%%%%%%%%%%%%%%%%%%%%%
The change of the thermodynamic ensemble corresponds to 
the addition of finite counterterms to the regularized Euclidean  action, which change the boundary conditions.  Let us illustrate this with a simple example. 

The Lorentzian action \eqref{Ireg1} is finite, and it has a well-defined action principle -- namely, it has an extremum when $E$, $Q$ and $a$ are kept fixed on the boundary. We can change these boundary conditions by adding one more finite counterterm to $I^{(a)}$,
\begin{equation}
    I_{\mathrm{fin}}^{(A)}=-\frac{1}{4}\,\Omega _{4}\int \mathrm{d}r\,\frac{\mathrm{d}}{\mathrm{d}r}\left[ \left( \Xi \varphi +\frac{2}{3}\,\varphi ^{3}-\phi +\frac{1}{3}\,\phi ^{3}\right) A_{t}\right] ,  \label{Ifin}
\end{equation}
to get the action
\begin{equation}
I^{(a,A)}=I^{(a)}+I_{\mathrm{fin}}^{(A)}\,.  \label{ct}
\end{equation}
The change $I^{(a)}\rightarrow I^{(a,A)}$ amounts to replacing
\begin{equation*}
    -\frac{1}{4}\,\left( \Xi \varphi +\frac{2}{3}\,\varphi ^{3}-\phi +\frac{1}{3}\,\phi ^{3}\right) 'A_{t}\rightarrow \frac{1}{4}\,\left( \Xi\varphi +\frac{2}{3}\,\varphi ^{3}-\phi +\frac{1}{3}\,\phi ^{3}\right) A'_{t}\,.
\end{equation*}
In its variation, the
replacement $\delta I^{(a)}\rightarrow \delta I^{(a,A)}$ corresponds to
\begin{equation*}
    -\Omega _{4}\,\frac{1}{4}\delta \left( \Xi _{0}C+\frac{2}{3}\,C^{3}-B+\frac{1}{3}\,B^{3}\right) \,A_{0}\rightarrow \Omega _{4}\,\frac{1}{4}\left( \Xi_{0}C+\frac{2}{3}\,C^{3}-B+\frac{1}{3}\,B^{3}\right) \delta A_{0}\,,
\end{equation*}
such that the total variation is
\begin{equation}
    \delta I^{(a,A)}=\Delta t\,({-}N_{0}\delta E+{Q\delta A_{0}}+P\delta a)\,.
    \label{varIaA}
\end{equation}
Thus, the Neumann boundary condition on $Q$ in $I^{(a)}$ is replaced by
the Dirichlet boundary condition on $A_{0}$ in $I^{(a,A)}$.

In the Euclidean space,  the addition of the boundary term (transformation \eqref{ct}) is equivalent to performing the Legendre transformation $I^{E(a)} \to I^{E(a,A)}$, which amounts to the change $A_0 \delta Q \to - Q\delta A_0$ in the variation of the thermodynamic potential. In consequence, the ensemble changes, as well, since the dependence on the dynamical variable $Q$ is replaced by the one on the dynamical variable $A_0$.

The Legendre transformation \eqref{G} obtained from the  partition function \eqref{partition}  and the Euclidean action is known as the quantum statistical relation \cite{Gibbons:2004ai}.
In particular, when no charges apart from the entropy are Legendre-transformed (i.e., 
$\bar a = \emptyset$, $\underline{a}=a$), the ensemble is canonical, and the corresponding free energy results in the Helmholtz free energy, depending only on the temperature and the chemical potentials. In the opposite extreme case, when all charges are Legendre-transformed (i.e., $\underline{a}=\emptyset$), the ensemble becomes grand canonical, with Gibbs free energy depending only on the temperature and the charges. We will call all the intermediate cases also grand canonical ensembles, where the free energy depends at least on one charge.

In the generic ensemble \eqref{G}, the Euclidean action is related to the thermodynamic quantities as
\begin{equation}
    I^E=\beta E-S-\beta\sum\limits_{{\bar a}} \mu_{\bar a} Q_{\bar a}\,. \label{Gcanonical}
\end{equation}
It is important to emphasize, however, that the quantum statistical relation \eqref{G} and the identification \eqref{G=TI}, are valid only in the (grand) canonical ensemble, in which the temperature is one of the independent variables, and not in the microcanonical ensemble, in which the energy is an independent variable. We will see in the next section that our treatment is naturally realized in the microcanonical ensemble. As a consequence, we shall not use our regularized action for the identification \eqref{G=TI}, but one related to it by Legendre transformation to (grand) canonical ensemble.

%%%%%%%%%%%%%%%%%%%%%%%
\subsection{Entropy}
%%%%%%%%%%%%%%%%%%%%%%%

We use the Lagrangian method in the minisuperspace approximation to compute the black hole entropy. Since the action $I^{(a)E}$ is already finite and satisfies a well-defined action principle in the Lorentzian space, no additional boundary terms, namely counterterms, are needed, except for finite counterterms when one wishes to change the ensemble.  Covariant methods to study black hole thermodynamics have also been employed in \cite{Blagojevic:2006nf,Gibbons:2004ai, Banados:1998ys,Miskovic:2010ey}. This should be contrasted with Hamiltonian approaches \cite{Regge:1974zd,Crisostomo:2000bb,Blagojevic:2019gsd,Blagojevic:2022etm}, where one adds a boundary term $B^{E}$, to the on-shell Euclidean action and chooses it so that the total action, $I^E=I^{(a)E}+B^E$, satisfies the variational principle, namely $\delta B^{E}=-\delta I^{(a)E}$. We will not follow that approach here.

In the Euclidean section, the spacetime region outside the black hole has two physical boundaries, infinity and the horizon, and the action principle must be satisfied at both.\footnote{This still leaves open the possibility of adding different boundary terms at the horizon. In \cite{Banados:1998ys}, in Einstein gravity, the horizon was not treated as a genuine boundary, and therefore no counterterms were introduced there.} This will be essential for deriving the first law of black hole thermodynamics.

The on-shell variation of the action \eqref{Ireg1} is a boundary term,
\begin{equation}
\delta I^{(a)E}=\beta \Omega _{3}\int\limits_{r_{+}}^{\infty }\mathrm{d}r\,
\frac{\mathrm{d}}{\mathrm{d}r}\,
\delta \Lambda ^{(a)}(r) \,,
\label{EuclidVar}
\end{equation}
where the variation of the integrand has three contributions, $\delta \Lambda ^{(a)}=\delta \mathring{\Lambda}+\delta \Lambda _{\mathrm{ct}}+\delta \Lambda _{\mathrm{fin}}^{(a)}$, obtained from the corresponding Lagrangian densities on-shell as $\delta\mathcal{L}=\beta \Omega_3\,\frac{\mathrm{d}(\delta\Lambda)}{\mathrm{d}r}$. The variations are taken in the fields $\Phi^\alpha=(N,f^{2},\psi ,\varphi ,\phi, A_{t})$, such that 
\begin{equation}
\delta \Lambda^{(a)} =\frac{\partial \Lambda^{(a)} }{\partial \Phi ^{\alpha }}\,\delta
\Phi ^{\alpha }+\frac{\partial \Lambda^{(a)} }{\partial \Phi ^{\prime \alpha }}
\,\delta \Phi ^{\prime \alpha }\,.
\end{equation}
The bulk variation contribution in \eqref{EuclidVar} is read off from \eqref{deltaI}
and the counterterms variation from \eqref{Ictfundnew}, namely
\begin{eqnarray}
\delta \mathring{\Lambda} &=&\left[ -\left( \frac{r^{3}}{3\ell ^{3}}-\frac{r\Xi }{%
2\ell }\right) \left( \rule[2pt]{0pt}{11pt}\delta (f(fN)^{\prime })+\delta
\psi \right) -\left( r\psi +rf(fN)^{\prime }-\frac{N\Xi }{2}\right) \frac{\,f\delta f}{\ell }\right.   \notag \\
&&-\frac{Nf^{2}}{\ell }\,\varphi \delta \varphi +\frac{1}{4}\,\delta \left(
\Xi \varphi +\frac{2}{3}\,\varphi ^{3}-\phi +\frac{1}{3}\,\phi ^{3}\right)
A_{t}\,, \label{bareLambda}
\end{eqnarray}
and
\begin{equation}
\delta \Lambda _{\mathrm{ct}}=\frac{r^{3}\delta \psi }{3\ell ^{3}}+r\delta
\left( \rule[2pt]{0pt}{11pt}f(fN)' \right) \left( \frac{r^{2}}{3\ell
^{3}}-\frac{\Xi }{2\ell }\right) -rf(fN)' \, \frac{\delta \Xi }{2\ell }
\,.
\end{equation}
The finite counterterm variation in \eqref{EuclidVar} is obtained from \eqref{Ifin1} as
\begin{equation}
\delta\Lambda_{\mathrm{fin}}^{(a)}=-\delta\left[ \frac{r\psi} {2\ell }\left( \frac{r^{2}}{\ell ^{2}}
-f^{2}\right) \right] \,.
\end{equation}
Summing up, all variations of the derivatives of the fields, $\delta \Phi ^{\prime \alpha }$, cancel out and
the total variation has a simple form 
\begin{eqnarray}
\delta \Lambda ^{(a)} &=&\frac{N\Xi }{4\ell }\,\delta f^{2}+\left( \Xi
+f^{2}-\frac{r^{2}}{\ell ^{2}}\right) \frac{r\delta \psi }{2\ell }+\left( 
\rule[2pt]{0pt}{12pt}rf(fN)^{\prime }-Nf^{2}\right) \frac{\varphi \delta
\varphi }{\ell }  \notag \\
&&+\frac{1}{4}\,\delta \left( \Xi \varphi +\frac{2}{3}\,\varphi ^{3}-\phi +
\frac{1}{3}\,\phi ^{3}\right) A_{t}\,.
\label{varLambda}
\end{eqnarray}
Evaluated at the asymptotic boundary, we get
\begin{eqnarray}
\left. \delta \Lambda ^{(a)}\right\vert _{\infty } &=&\frac{N_{0}\Xi _{0}}{%
4\ell }\,\delta (-e^{2})+\left( \Xi _{0}-e^{2}\right) \frac{\delta a}{2\ell }%
+e^{2}N_{0}\,\frac{C\delta C}{\ell }  \notag \\
&&+\frac{1}{4}\,\delta \left( \Xi _{0}C+\frac{2}{3}\,C^{3}-B+\frac{1}{3}
\,B^{3}\right) A_{0}\,.
\end{eqnarray}
In the first line, we recognize $\Omega _{3}^{-1}\left( -N_{0}\delta E+P\delta a\right) 
$ from \eqref{E} and \eqref{Pmini} while, in the second line, the variation of the charge $Q$ appears in the form $-\Omega _{3}^{-1}\delta Q\,A_{0}$, as it can be seen from \eqref{Qmini}.  Therefore, this expression evaluated on-shell at $r\rightarrow \infty $ leads to the known result from the Lorentzian analysis,
\begin{equation}
\left. \delta \Lambda ^{(a)}\right\vert _{\infty }=\Omega _{3}^{-1}\left(\rule[2pt]{0pt}{10pt} -N_{0}\delta E+P\delta a-A_{0}\delta Q\right) \,.\label{Lambda.infty}
\end{equation}

On the other hand, to evaluate \eqref{varLambda} at the horizon, we have to vary the fields
keeping $\delta r=0$, and we find 
\begin{eqnarray}
\delta \Lambda ^{(a)} &=&\left( \frac{N\Xi }{4\ell }-\frac{1}{4}%
\,A_{t}\,\varphi \right) \,\delta f^{2}+\left( 1-\varphi ^{2}\right) \,\frac{%
r\delta \psi }{2\ell }+\frac{1}{4}\,A_{t}\left( -1+\phi ^{2}\right) \delta
\phi   \notag \\
&&+\left[ \left( \rule[2pt]{0pt}{12pt}rf(fN)^{\prime }-Nf^{2}\right) \varphi
+\frac{1}{4}\,\Xi \,\ell A_{t}\right] \frac{\delta \varphi }{\ell }\,.
\end{eqnarray}
In our ensemble, the three independent variables are $r_{+}=\ell e$, $a$ and $C$. Varying in them while $
\delta N_{0}=0$, $\delta A_{0}=0$ and $\delta B=0$, and then evaluating the
result at $r_{+}$, the non-vanishing field variations are
\begin{eqnarray}
\left.\delta f^{2}\right\vert
_{r_+}=-2e\delta e\,, 
\qquad \left. \delta \psi \right\vert
_{r_+} = \frac{\delta a}{\ell e}\,,  \qquad
\left. \delta \varphi \right\vert
_{r_+} =\delta C\,,
\label{varPhi(r+)}
\end{eqnarray}
while $\delta \phi |_{r_{+}}$ vanishes. 
Replacing also $N'|_{r_+}=0$, $ff'|_{r_{+}}=\frac{e}{\ell }$, and $\Xi _{0}$ from \eqref{Xi-charge}, we get
\begin{eqnarray}    
\delta \Lambda^{(a)}|_{r_{+}}&=&\left[ \frac{N_{0}}{2}\,\left(
e^{2}-C^{2}+1\right)  -
\frac{\ell C}{2}\,A_{0}\right] \frac{e\delta e}{\ell } + \frac{1-C^2}{2\ell}\, \delta a  \notag \\
&& +\left[ (e^2+1-C^2) \, \frac{A_0}{4}+\frac{e^2N_0C}{\ell} \right] \delta C\,.
\end{eqnarray}

The variation of the Euclidean on-shell action \eqref{EuclidVar} has the form
\begin{equation}
 \delta I^{(a)E} =\beta \Omega _{3}\left( \rule[2pt]{0pt}{10pt}\delta
\Lambda ^{(a)}|_\infty-\delta \Lambda ^{(a)}|_{r_{+}}\right).
\end{equation}
As we saw in \eqref{Lambda.infty}, the first term is related to the asymptotic charges and chemical potentials, while we denote the second term, the horizon contribution, by $\delta\mathscr{S}$, namely,
\begin{equation}
\delta I^{(a)E}=\beta\left( -N_{0}\delta E+P\delta a-A_{0}\delta
Q\right) + \delta \mathscr{S}\,,
\label{dI(a)E}
\end{equation}
where
\begin{equation}
\delta \mathscr{S}=-\beta \Omega _{3}\left. \delta \Lambda ^{(a)}\right\vert _{r_{+}}\,,
\end{equation}
or explicitly 
\begin{eqnarray}
\delta \mathscr{S} &=&\beta \Omega _{3} \left\{ \left[ N_{0}\,\left(
e^{2}-C^{2}+1\right)  -
\ell C\,A_{0}\right] \frac{e\delta e}{2\ell }  - \frac{1-C^2}{2\ell}\, \delta a   \right. \notag \\
&& - \left. \left[ (e^2-C^2+1) \, \frac{A_0}{4}+\frac{e^2N_0C}{\ell} \right] \delta C \right\}
\,.
\label{dS}
\end{eqnarray}
Imposing stationarity of the Euclidean on-shell action \eqref{dI(a)E}, namely
$\delta I^{(a)E}=0$, leads to the first law of thermodynamics once the quantity
$\mathscr{S}$ is identified with the entropy,
\begin{equation}
    S=\mathscr{S}\,.
    \label{S}
\end{equation}
Multiplying the resulting relation by $T$, the first law takes the form
\begin{equation}
\label{1stlaw explicit}
\delta E=T\delta S+P\delta a-A_{0}\delta Q\,. 
\end{equation}
Accordingly, the first law is expressed in the form \eqref{1st law}, in terms of the internal energy $E=E(S,a,Q)$, as appropriate to a microcanonical description, rather than as a free energy.

As already mentioned, the Euclidean action $I^{(a)E}$ cannot be identified with the internal energy, defined in the microcanonical ensemble.  Indeed, an explicit evaluation of our regularized Euclidean action, given by \eqref{Ireg0}, \eqref{Imini}, and \eqref{Ictfundnew}, together with
the finite counterterm \eqref{Ifin}, shows that the total Euclidean action
vanishes,
\begin{equation}
    I^E_{\mathrm{reg}}=0\,,\quad  I^{(a)E}_{\mathrm{fin}}=0 \quad \Rightarrow \quad I^{(a)E}=I^E_{\mathrm{reg}}+I^{(a)E}_{\mathrm{fin}}=0\,. 
\end{equation}
Passing to the canonical ensemble requires a Legendre transformation with respect to the entropy, which defines the Helmholtz free
energy, $G=E-TS=G(T,a,Q)$. In the new ensemble, the first law then takes the form \eqref{1st law},
\begin{equation}
\delta G=-S\delta T+P\delta a-A_{0}\delta Q\,.
\end{equation}

To be able to integrate it and compute the entropy $S$, its variation $\delta S =S_{e}\,\delta e+S_{a}\,\delta a+S_{C}\,\delta C$ has to satisfy the Frobenius
integrability conditions
\begin{equation}\label{var_S}
\frac{\partial S_{e}}{\partial a}=\frac{\partial S_{a}}{\partial e}\,,\qquad \frac{\partial S_{e}}{\partial C}=\frac{\partial S_{C}}{\partial e},\qquad \frac{\partial S_{a}}{\partial C}=\frac{\partial S_{C}}{\partial a}\,.
\end{equation}
Using  $\beta =\frac{2\pi \ell }{e}$ and $N_0=1$, it can be shown that the entropy variation \eqref{dS} does not fulfill \eqref{var_S}, thus the entropy is not integrable. This was expected, as the contribution in $\delta C$ was not integrable in the computation of the energy either  (see eq.~\eqref{E}). This problem can be circumvented naturally, if we recall that the parameter $C$ is not dynamical, $\delta C=0$, since it is a function of the discrete topological number due to eq.~\eqref{n12}. Using this fact, the entropy becomes integrable, and we get 
\begin{eqnarray}
S &=&\pi \Omega _{3}\int \biggl[\mathrm{d}e\,\left( 
e^{2}-C^{2}+1 -\ell C\,A_{0}\right) + \frac{C^2-1}{2\ell}\, \mathrm{d}a\biggr]  \notag \\
&=&\pi \Omega _{3}\,\left[ \left( \frac{e^{2}}{3}-C^{2}+1-\ell CA_{0}\right) e+ \frac{C^2-1}{2\ell}\,a\right].
\label{EntropyCS}
\end{eqnarray}
The integration constant is fixed so that the extremal BPS black hole, $e=0$, $B=C$, $a=0$ \cite{Andrianopoli:2021qli}, has zero entropy. The expression for $S(e,a)$ depends linearly on $a$, and therefore the entropy is not bounded from below unless the allowed values of $a$ are restricted. For fixed $C$ this happens, for instance, if the geometric origin of the trace torsion enforces a definite sign for $(C^2-1)a$. In such cases, the integration constant can be chosen so that $S$ is non-negative, and the extremal black hole has non-vanishing entropy.\footnote{This is what happens in extended supergravity, where the black-hole entropy is proportional to the area of the event horizon \cite{Kallosh:1992ii,Kallosh:1992wa} and, 
in the extremal limit, it becomes a topological quantity, completely fixed in terms of the quantized charges \cite{Ferrara:1995ih,Ferrara:1997tw}. A microscopic statistical interpretation of the entropy in these theories, related to microstate counting, was found in non-perturbative string theory
\cite{Strominger:1996sh}. } A more careful analysis of the geometric interpretation of the constant $a$ in the full CS AdS gravity theory, rather than in its reduced version, is needed to determine whether $a$ can indeed take arbitrary values.

In the torsionless case without additional charges ($C=0$, $a=0$, $A_{0}=0$), the entropy reduces to
\begin{equation}
S_{\mathrm{torsionless}}=\pi \Omega _{3}\left( \frac{e^{3}}{3}+e\right) .
\end{equation}
This result matches the one in \cite{Crisostomo:2000bb} that is $S_{\mathrm{BHscan}}=\frac{4\pi }{G_{2}}\left( r_{+}+\frac{r_{+}^{3}}{3\ell ^{2}}\right)$, if we use footnote \ref{BH scan} to  compare the notations.

%%%%%%%%%%%%%%%%%%%%%%%%%%%%%
\section{Other formulas for non-Riemannian entropy}
\label{other}
%%%%%%%%%%%%%%%%%%%%%%%%%%%%%

In the previous section, we computed the black hole entropy from the action principle using the minisuperspace approximation of CS AdS supergravity and found an expression consistent with the first law of thermodynamics. To further test this result, we apply known entropy formulas in Riemann–Cartan spacetime derived from the full CS supergravity theory, which do not rely on the minisuperspace approximation. Consequently, the parameter 
$a$ does not enter the computation, and in this section we set $a=0$.

%%%%%%%%%%%%%%%%%%%%%%%%%%%%%%
\subsection{Wald's formula}
%%%%%%%%%%%%%%%%%%%%%%%%%%%%%

We compute entropy from Wald’s formula \cite{Wald:1993nt} generalized to spacetimes with torsion in \cite{Gallegos:2020otk}. To make a comparison, we set $A_0=0$ and  $B=0$, that is, the axial torsion $C$ is the only new parameter. The entropy is given by the formula 
\begin{equation}    
S=\frac{\pi k}{\ell}\int\limits_{\Sigma_+}\epsilon_{abcde}\;n^{ab}\left(R^{cd}e^e+\frac{1}{3\ell ^2}\,e^c e^d e^e   \right),
\label{SWald}
\end{equation} 
where $n^{ab}=D^{[a}\zeta^{b]}|_{\Sigma_+}= g^{\mu\nu}e^{[a}_{\;\mu}\Diff_{\nu}\zeta^{b]}|_{\Sigma_+}$ is a binormal defined as an antisymmetrized covariant derivative of the Killing vector field $\zeta=\zeta^{\mu} \partial _\mu$ generating the horizon $\Sigma_+$ for the Lorentz-covariant derivative $D_\mu (\omega)$.  Note that the entropy is dimensionless as there is the factor $1/\ell$ by each vielbein, which is the only dimensionful quantity in \eqref{SWald}.

For a static black hole, the Killing vector field is $\partial_t$. However, the formalism requires that the vector field be normalized as $\zeta=\kappa^{-1} \, \partial_t$, where $\kappa=2\pi T$ is the surface gravity. For the temperature given by \eqref{T}, with $N_0=1$, it gives $\zeta=\frac{ \ell^2}{r_+}\, \partial_t$. 

The only non-vanishing component of the binormal is $n^{01}=-n^{10}=\frac{1}{2}$, which has the same form as in the Riemannian spacetime. This follows from  
\begin{eqnarray}
 \Diff^0\zeta^1=\frac{r}{r_+}\,, \qquad \Diff^1\zeta^0=0\,.
\end{eqnarray}

With these results at hand,  we project \eqref{SWald} on the horizon,
\begin{equation}
S=\frac{\pi k}{\ell} \int\limits_{\Sigma_+}\epsilon _{ijk}\left( R^{ij}e^{k}+\frac{1}{3\ell ^{2}}\,e^{i}e^{j}e^{k}\right) \,,
\end{equation}
and evaluate it on the solution \eqref{e,w}, \eqref{bh2}, taking into account that $f^2(r_+)=0$. The result is
\begin{eqnarray}
S &=&\frac{\pi k}{\ell} \int\limits_{\Sigma_+}\epsilon _{ijk}\,\tilde{e}^{i}\tilde{e}^{j}\tilde{e}^{k}\left( r_{+}(1-\varphi ^{2})+\frac{r_{+}^{3}}{3\ell ^{2}}\right) \,,
\end{eqnarray}
or setting $e=\frac{r_+}{\ell}$ and after integration,
\begin{eqnarray}
S &=&\pi k \,3!\,\mathrm{Vol}(\Sigma )\,\left( \frac{e^2}{3}+1-C^2\right)e\,.
\end{eqnarray}
Since $\Omega_3=6 k \mathrm{Vol}(\Sigma )$, the final result
matches eq.~\eqref{EntropyCS} when $a=0$ and $A_0=0$. 

Therefore, our results are consistent with the Wald's entropy of a five-dimensional black hole with torsion in CS AdS supergravity. The entropy formula found in \cite{Gallegos:2020otk} takes the same form as Wald’s formula in the Riemannian case, with the Riemann curvature replaced by the Riemann--Cartan one. Similarly, holographic arguments \cite{Banados:2006fe} (see also section \ref{vacuum-energy-section}) suggest that the Weyl anomaly in Riemann-Cartan space preserves its Riemannian form under the same substitution. This raises the interesting question of whether the holographic entanglement entropy in this model may also be obtained in an analogous manner, as conjectured in \cite{Dordevic:2026pct}.

%%%%%%%%%%%%%%%%%%%%%%%%%%%%%%
\subsection{Hamiltonian  entropy}
%%%%%%%%%%%%%%%%%%%%%%%%%%%%%

Entropy can be computed using the Hamiltonian formalism of \cite{Blagojevic:2019gsd}, that can include also gauge fields \cite{Blagojevic:2022etm}. In this framework, the charges are obtained through the Regge–Teitelboim procedure \cite{Regge:1974zd}, namely as boundary terms that render the canonical symmetry generator differentiable. An advantage of this approach is that it applies to all fields entering the generator, including the torsion field and additional gauge fields.

To summarize the method, denote the covariant momenta 3-forms as
\begin{eqnarray}
    \Pi_{ab}=\frac{\partial L}{\partial R^{ab}}\;,\qquad \Pi_a=\frac{\partial L}{\partial T^a}\;, \qquad \Pi=\frac{\partial L}{\partial F}\,, \qquad \widetilde{\Pi}_i=\frac{\partial L}{\partial \mathcal{F}^i}\,,
\end{eqnarray}
where $L=k(L_{\mathrm{G}}+L_{\mathrm{SU}(2)}+L_{\mathrm{int}})$ is the Lagrangian 5-form read off from \eqref{action} and \eqref{L}. Then the variation of the charge density 3-form is given by \cite{Blagojevic:2019gsd},
\begin{eqnarray}
\delta \mathcal{G} &=&\iota _{\xi }e^{a}\delta \Pi _{a}+\delta e^{a}\iota
_{\xi }\Pi _{a}+\frac{1}{2}\,\iota _{\xi }\omega ^{ab}\delta \Pi _{ab}+\frac{1}{2}\, \delta \omega ^{ab}\iota _{\xi }\Pi _{ab}  \notag \\
&&+\iota _{\xi }A\, \delta \Pi +\delta A\, \iota _{\xi }\Pi +\iota _{\xi
}\mathcal{A}^{I}\delta \widetilde{\Pi} _{I}+\delta \mathcal{A}^i\iota
_{\xi }\widetilde{\Pi}_i\;,  \label{var.charge}
\end{eqnarray}
where the variations are taken in the independent thermodynamic parameters and 
the contraction $\iota_{\xi}$ is performed by the Killing vector field $\xi=\partial_t$. Boundary charges (energy and $\mathrm{U}(1)$ charge) are obtained from the boundary integral $\delta \Gamma (\Sigma_\infty)= \int_{\Sigma_\infty} \delta \mathcal{G}$, where the integration is performed over the surface $\Sigma \simeq \mathbb{S}^3$ located at the infinite radius, while the entropy corresponds to the horizon contribution, 
\begin{equation}
\delta\Gamma(\Sigma_+)=\int\limits_{\Sigma_+} \delta \mathcal{G}=T\delta S\,.
\label{S.ham}
\end{equation}
The first law of thermodynamics is implied from the conservation of these quantities, namely,
\begin{equation}
   \delta \Gamma(\Sigma_\infty)=\delta\Gamma(\Sigma_+)\;.
\end{equation}
Therefore, to obtain the entropy, we first define the momenta, 
\begin{eqnarray}
    \Pi_{ab} &=& \frac{k}{2\ell}\, \epsilon_{abcde}\, \left( R^{cd}+\frac{1}{3\ell^2}\, e^ce^d \right)e^e+\frac{k}{4}\left(R_{ab}+\frac{1}{\ell^2}\, e_ae_b  \right)A\,,\notag \\
   \Pi_a &=&-\frac{k}{2\ell^2}\, T_a A\,,\qquad  \Pi=0\,,\qquad \widetilde{\Pi} _i= -\frac{k}{2}\,\mathcal{F}_i\,A\,,
\end{eqnarray}
and then we evaluate it in terms of $\Phi^\alpha$ using eqs.~\eqref{e,w}--\eqref{T-ans} and \eqref{ans1}--\eqref{ans}. Furthermore, the terms along $\diff r$ do not contribute at $\Sigma$ defined at $r=\mathrm{const}$, such that we obtain the following non-vanishing components of the momenta with the indices $a=(0,1,i)$:
\begin{eqnarray}
\Pi_{01} &=&\frac{r\Omega _{3}}{2\ell }\,\left( 1-\varphi ^{2}-f^{2}+\frac{
r^2}{3\ell ^{2}}\right) \,,  \notag \\
\Pi_{1i} &=&\frac{k}{2\ell }\left[ -2r\sigma +\left( 1-\varphi ^{2}-f^{2}+
\frac{r^{2}}{\ell ^{2}}\right) N-\ell \varphi \,A_{t}\right] f\epsilon
_{ijk}\,\mathrm{d}t\,\tilde{e}^{j}\tilde{e}^{k}\,, \notag \\
\Pi_{ij} &=&\frac{k}{2}\left[ -\frac{4}{\ell }\,Nf^{2}\varphi +\left(
1-\varphi ^{2}-f^{2}+\frac{r^{2}}{\ell ^{2}}\right) A_{t}\right] \mathrm{d}%
t\,\tilde{e}_{i}\tilde{e}_{j}\notag
\end{eqnarray}
 and 
\begin{eqnarray}
\Pi_{i} &=& -\frac{k}{2\ell ^{2}}\,r\varphi A_{t}\,\epsilon _{ijk}\,\mathrm{d}t \,\tilde{e}^{j}\tilde{e}^{k}\,,\notag\\
\widetilde{\Pi }_i&=&-\frac{k}{4}\,\left( 1-\phi^{2}\right) \,A_{t}\,\epsilon _{ijk}\,\tilde{e}^{j}\tilde{e}^{k}\mathrm{d}t \,.
\end{eqnarray}

 As regards the contractions of the fields, the non-vanishing components are
\begin{equation}
\iota _{\xi }e^{0}=Nf\,,\qquad \iota _{\xi }\omega ^{01}=-\iota _{\xi
}\omega ^{10}=\frac{Nr}{\ell ^{2}} \to \sigma\,, \qquad \iota _{\xi } A=A_t\,.
\end{equation}

With these results at hand, the variation of the charge density that
contains only non-vanishing terms is 
\begin{equation}
\delta \mathcal{G}=\delta e^{i}\iota _{\xi }\Pi _{i}+\iota _{\xi }\omega
^{01}\delta \Pi _{01}+\omega ^{1i}\iota _{\xi }\Pi _{1i}+\frac{1}{2}\,\delta
\omega ^{ij}\iota _{\xi }\Pi _{ij}+\delta \mathcal{A}^i\,\iota _{\xi }
\widetilde{\Pi }_i\,.
\end{equation}%
In terms of the fields, it reads 
\begin{eqnarray}
\delta \mathcal{G} &=&-\frac{k}{2\ell ^{2}}\,r\varphi A_{t}\,\epsilon
_{ijk}\,\delta \left( r\,\tilde{e}^{i}\right) \tilde{e}^{j}\tilde{e}^{k}+%
\frac{k}{2\ell }\,\sigma \,\delta \left[ r\left( 1-\varphi ^{2}-f^{2}+\frac{%
r^{2}}{3\ell ^{2}}\right) \right] \epsilon _{ijk}\tilde{e}^{i}\,\tilde{e}^{j}%
\tilde{e}^{k}  \notag \\
&&-\frac{k}{2\ell }\left[ -2r\sigma +\left( 1-\varphi ^{2}-f^{2}+\frac{r^{2}%
}{\ell ^{2}}\right) N-\ell \varphi \,A_{t}\right] \epsilon _{ijk}\,f\delta
(f\,\tilde{e}^{i})\tilde{e}^{j}\tilde{e}^{k}  \notag \\
&&+\frac{k}{4}\,\epsilon ^{ijk}\left[ -\frac{4}{\ell }\,Nf^{2}\varphi
+\left( 1-\varphi ^{2}-f^{2}+\frac{r^{2}}{\ell ^{2}}\right) A_{t}\right] \,%
\tilde{e}_{i}\tilde{e}_{j}\delta \left( \tilde{\omega}_{k}-\varphi \,\tilde{e%
}_{k}\right) \notag \\
&&-\frac{k}{4}\,\left(
1-\phi ^{2}\right) \,A_{t}\,\epsilon _{ijk}\,\tilde{e}^{j}\tilde{e}^{k}\,\delta (\tilde{\omega}^{i}-\phi \,\tilde{e}^{i})\,.
\end{eqnarray}
The entropy is obtained from the on-shell evaluation of the charge density $\delta \mathcal{G}$ at the horizon, using the solution \eqref{bh2} with $a=0$, 
$N_{0}=1$ and only $\delta e\neq 0$, while $\delta \tilde{e}^{i}=0$ and $\delta \tilde{\omega}^{i}=0$. Namely, from the
three variables $e,C,a$, only $e$ varies non trivially. Also, $r_{+}=\ell e$
and $\delta r_{+}=\ell \,\delta e\neq 0$, such that, using also $f^2(r_{+})=0$
and $f(r_{+})\delta f(r_{+})=0$, the variation of the entropy density acquires the
form
\begin{equation}
\delta \mathcal{G}|_{r=r_{+}}=\frac{k}{4\ell }\,\left( 1-C^{2}+e^{2}-\ell
CA_{0}\right) \delta e^{2}\epsilon _{ijk}\tilde{e}^{i}\,\tilde{e}^{j}\tilde{e}^{k}\,.
\end{equation}
It can be seen that the $\mathrm{SU}(2)$ contribution drops out because $\delta B=0$.

Applying $\Omega _{3}=k\,\int_{\Sigma_+}\epsilon _{ijk}\tilde{e}^{i}\,
\tilde{e}^{j}\tilde{e}^{k}$ and the formula \eqref{S.ham}, the entropy variation is identified from
\begin{equation}
T\delta S=\int \delta \mathcal{G}|_{r=r_{+}}=\frac{e}{2\ell }\,\Omega
_{3}\left( 1-C^{2}+e^{2}-\ell CA_{0}\right) \delta e\,.
\end{equation}
It can be integrated out by keeping $A_{0}$ fixed, to get 
\begin{equation}
T\delta S=\frac{e}{2\ell }\,\Omega _{3}\delta \left[ \left( \frac{e^{2}}{3}%
+1-C^{2}\right) e-\ell CA_{0}e\right] \,.
\end{equation}%
Taking into account that $T=\frac{e}{2\pi \ell }$ as given by \eqref{T}, we obtain that the final expression for the entropy matches the result \eqref{EntropyCS} when $a=0$.

%%%%%%%%%%%%%%%%%%%%%%%%%%%%%
\section{Conclusions}
\label{conclusions}
%%%%%%%%%%%%%%%%%%%%%%%%%%%%%

We have shown that the thermodynamics of the five-dimensional Chern--Simons
AdS black hole coupled to $\mathrm{SU}(2)$ solitons \cite{Andrianopoli:2021qli}
can be consistently described within a minisuperspace approximation that preserves
the relevant dynamics of the $\Xi _{0}$-charged sector. The integration
constants characterizing the geometry are the extremality parameter $%
e^{2}\geq 0$, related to the black hole mass, the soliton magnitudes $B$ and 
$C$, the $\mathrm{U}(1)$ potential $A_{0}$, and the lapse function $N_{0}$,
which is set to $1$ after the boundary conditions are applied. The reduced
action reproduces the known black hole branch, augmented by an additional
parameter $a$ associated with a new torsional degree of freedom $T^0_{tr}$, which naturally appears from the integration of the field equations in the minisuperspace approximation, and
is needed to form three pairs of conjugate variables. The method provides a
transparent variational framework in which the boundary terms determine
three conserved quantities and their conjugate variables. In particular, one
recovers the variation of the energy and the $\mathrm{U}(1)$ charge
previously obtained by Hamiltonian methods \cite{Andrianopoli:2021qli}, while the enlarged
parameter space also reveals a momentum conjugate to the trace-torsion mode.

Within this framework, the Euclidean action leads to an entropy satisfying
the first law, and therefore to a consistent thermodynamic interpretation of
the solution. A central result is that the axial torsion parameter $C$
contributes nontrivially to the entropy, unlike what happens in many other
torsional black hole models. At the same time, when the topological sector
is fixed and $\delta C=0$, since $C$ is also a quantized topological charge,
the energy becomes integrable. In this sense, $C$ is naturally interpreted
as a conserved, but not primary, black hole hair -- it modifies the
thermodynamic state, but does not enter the first law through an independent
variation.

It is interesting to note that $C$ may be viewed as analogous to the
magnetic mass in Einstein--AdS gravity, as it is a topological property of
the geometry, characterized by the Pontryagin topological invariant, which
modifies the Noether charge \cite{Araneda:2016iiy}.

Finally, the entropy obtained from the reduced action is confirmed by two
independent methods in the full theory: the generalized Wald formula \cite{Wald:1993nt,Gallegos:2020otk}
and the Hamiltonian approach in Riemann--Cartan spacetime \cite{Blagojevic:2019gsd,Blagojevic:2022etm}. Their agreement provides a nontrivial consistency check of the analysis and
supports the conclusion that five-dimensional Chern--Simons AdS gravity
offers a genuine example in which torsion plays an essential role in
black hole entropy. This suggests interesting directions for future work,
especially concerning the physical interpretation of the extended parameter
space and the holographic role of torsion.

Another open question worth exploring is the role of the integration constant $a$ in the full Chern--Simons AdS gravity theory, where it is associated with the trace component of the torsion field.

It would also be interesting to study other branches of black hole solutions admitting the BPS states discussed in \cite{Andrianopoli:2021qli}, and to determine whether the minisuperspace approximation provides a consistent truncation for them as well. One such branch is the so-called $\Xi _{0}$-neutral solution, in which both the internal $\mathrm{SU}(2)$ gauge field and the AdS-$\mathrm{SU}(2)$  gauge field vanish at the asymptotic boundary.

Possible extensions of this work include the analysis of the Einstein--Gauss--Bonnet AdS black hole with torsion found in \cite{Banados:2001hm}, as well as exploring whether the present method can be
applied to the solution of \cite{Miskovic:2006ei}, which is not a black hole. A further direction is to explore a holographic correspondence between the model presented here and condensed matter systems, focusing in particular on the boundary quantum effects associated with bulk torsion components.

In summary, our results indicate that the minisuperspace approach captures
the essential thermodynamic features of this class of torsional solutions
and provides a useful framework for further investigations of torsion in
Chern--Simons AdS gravity.

%%%%%%%%%%%%%%%%%%%%%%%%%%%%%
\section*{Acknowledgments}
%%%%%%%%%%%%%%%%%%%%%%%%%%%%%

We are grateful to Ruben Monten and Jorge Zanelli for their inspiring remarks on black hole entropy. This work was supported in part by FONDECYT Regular Grants No.~1230492 and 1231779.
D.D.~was supported by the Science Fund of the Republic of Serbia under grant TF C1389-YF,  {\it Towards a Holographic Description of Noncommutative Spacetime: Insights from Chern-Simons Gravity, Black Holes and Quantum Information Theory} (HINT). O.M.~also gratefully acknowledges the hospitality of the Faculty of Physics, University of Belgrade, during a one-month research visit hosted by HINT.

%%%%%%%%%%%%%%%%%%%%%%%%%%%%%%%%%%%%%%%%%%


\begin{thebibliography}{9}

\bibitem{Chamseddine:1976bf}
A.~H.~Chamseddine and P.~C.~West,
``Supergravity as a Gauge Theory of Supersymmetry,''
Nucl. Phys. B \textbf{129} (1977), 39-44.

\bibitem{Troncoso:1998ng}
R.~Troncoso and J.~Zanelli,
``Gauge supergravities for all odd dimensions,''
Int. J. Theor. Phys. \textbf{38} (1999), 1181-1206
[arXiv:hep-th/9807029 [hep-th]].

\bibitem{Zanelli:2005sa}
J.~Zanelli,
``Lecture notes on Chern-Simons (super-)gravities. Second edition (February 2008),''
[arXiv:hep-th/0502193 [hep-th]].

\bibitem{Banados:1996yj}
M.~Banados, L.~J.~Garay and M.~Henneaux,
``The Dynamical structure of higher dimensional Chern-Simons theory,'' Nucl. Phys. B \textbf{476} (1996), 611-635
[arXiv:hep-th/9605159 [hep-th]].

\bibitem{Miskovic:2003ex}
O.~Miskovic and J.~Zanelli,
``Dynamical structure of irregular constrained systems,''
J. Math. Phys. \textbf{44} (2003), 3876-3887
[arXiv:hep-th/0302033 [hep-th]].

\bibitem{Banados:1993ur}
M.~Banados, C.~Teitelboim and J.~Zanelli,
``Dimensionally continued black holes,''
Phys. Rev. D \textbf{49} (1994), 975-986
[arXiv:gr-qc/9307033 [gr-qc]].


\bibitem{Aros:2002rk}
R.~Aros, C.~Martinez, R.~Troncoso and J.~Zanelli,
``Supersymmetry of gravitational ground states,''
JHEP \textbf{05} (2002), 020
[arXiv:hep-th/0204029 [hep-th]].

\bibitem{Zegers:2005vx}
R.~Zegers,
``Birkhoff's theorem in Lovelock gravity,''
J. Math. Phys. \textbf{46} (2005), 072502
[arXiv:gr-qc/0505016 [gr-qc]].


\bibitem{Deser:2005gr}
S.~Deser and J.~Franklin,
``Birkhoff for Lovelock redux,''
Class. Quant. Grav. \textbf{22} (2005), L103-L106
[arXiv:gr-qc/0506014 [gr-qc]].


\bibitem{Canfora:2007xs}
F.~Canfora, A.~Giacomini and R.~Troncoso,
``Black holes, parallelizable horizons and half-BPS states for the Einstein-Gauss-Bonnet theory in five dimensions,''
Phys. Rev. D \textbf{77} (2008), 024002
[arXiv:0707.1056 [hep-th]].

\bibitem{Brihaye:2013vsa}
Y.~Brihaye and E.~Radu,
``Black hole solutions in d=5 Chern-Simons gravity,''
JHEP \textbf{11} (2013), 049
[arXiv:1305.3531 [gr-qc]].

\bibitem{Giribet:2014hpa}
G.~Giribet, N.~Merino, O.~Miskovic and J.~Zanelli,
``Black hole solutions in Chern-Simons AdS supergravity,''
JHEP \textbf{08} (2014), 083
[arXiv:1406.3096 [hep-th]].

\bibitem{Andrianopoli:2021qli} L.~Andrianopoli, G.~Giribet, D.~L.~D\'{\i}{}az and O.~Miskovic, ``Black holes with topological charges
in Chern-Simons AdS$_{5}$ supergravity,'' 
JHEP \textbf{11} (2021), 123 [arXiv:2106.01876 [hep-th]].

\bibitem{Miskovic:2006ei}
O.~Miskovic, R.~Troncoso and J.~Zanelli,
``Dynamics and BPS states of AdS(5) supergravity with a Gauss-Bonnet term,''
Phys. Lett. B \textbf{637} (2006), 317-325
[arXiv:hep-th/0603183 [hep-th]].

\bibitem{Edelstein:2010sx}
J.~D.~Edelstein, A.~Garbarz, O.~Miskovic and J.~Zanelli,
``Stable p-branes in Chern-Simons AdS supergravities,''
Phys. Rev. D \textbf{82} (2010), 044053
[arXiv:1006.3753 [hep-th]].

\bibitem{Liberati:2015xcp}
S.~Liberati and C.~Pacilio,
``Smarr Formula for Lovelock Black Holes: a Lagrangian approach,''
Phys. Rev. D \textbf{93} (2016) no.8, 084044
[arXiv:1511.05446 [gr-qc]].


\bibitem{Blagojevic:2006jk}
M.~Blagojevic and B.~Cvetkovic,
``Black hole entropy in 3-D gravity with torsion,''
Class. Quant. Grav. \textbf{23} (2006), 4781
[arXiv:gr-qc/0601006 [gr-qc]].

\bibitem{Blagojevic:2006nf}
M.~Blagojevic and B.~Cvetkovic,
``Covariant description of the black hole entropy in 3D gravity,''
Class. Quant. Grav. \textbf{24} (2007), 129-140
[arXiv:gr-qc/0607026 [gr-qc]].


\bibitem{Ma:2013eaa}
M.~S.~Ma and R.~Zhao,
``Phase transition and entropy spectrum of the BTZ black hole with torsion,''
Phys. Rev. D \textbf{89} (2014) no.4, 044005
[arXiv:1310.1491 [gr-qc]].


\bibitem{Klemm:2007yu}
D.~Klemm and G.~Tagliabue,
``The CFT dual of AdS gravity with torsion,''
Class. Quant. Grav. \textbf{25} (2008), 035011
[arXiv:0705.3320 [hep-th]].


\bibitem{Aviles:2023igk}
L.~Avil\'es, D.~Hidalgo and O.~Valdivia,
``Thermodynamics of the three-dimensional black hole with torsion,''
JHEP \textbf{09} (2023), 185
[arXiv:2308.09114 [gr-qc]].

\bibitem{Blagojevic:2019gsd}
M.~Blagojevi\'c and B.~Cvetkovi\'c,
``Entropy in Poincar\'e gauge theory: Hamiltonian approach,''
Phys. Rev. D \textbf{99} (2019) no.10, 104058
[arXiv:1903.02263 [gr-qc]].


\bibitem{Blagojevic:2022etm}
M.~Blagojevi\'c and B.~Cvetkovi\'c,
``Entropy of Kerr-Newman-AdS black holes with torsion,''
Phys. Rev. D \textbf{105} (2022) no.10, 104014
[arXiv:2203.14696 [gr-qc]].


\bibitem{Chakraborty:2018qew}
S.~Chakraborty and R.~Dey,
``Noether Current, Black Hole Entropy and Spacetime Torsion,''
Phys. Lett. B \textbf{786} (2018), 432-441
[arXiv:1806.05840 [gr-qc]].

\bibitem{Cvetkovic:2017nkg}
B.~Cvetkovi{\'c} and D.~Simi{\'c},
``A black hole with torsion in 5D Lovelock gravity,''
Class. Quant. Grav. \textbf{35} (2018) no.5, 055005
[arXiv:1707.01258 [gr-qc]].

\bibitem{Rogatko:2006hck}
M.~Rogatko,
``First Law of Black Rings Thermodynamics in Higher Dimensional Chern-Simons Gravity,''
Phys. Rev. D \textbf{75} (2007), 024008
[arXiv:hep-th/0611260 [hep-th]].


\bibitem{Correa:2013bza}
F.~Correa and M.~Hassaine,
``Thermodynamics of Lovelock black holes with a nonminimal scalar field,''
JHEP \textbf{02} (2014), 014
[arXiv:1312.4516 [hep-th]].


\bibitem{Bravo-Gaete:2021hza}
M.~Bravo-Gaete, C.~G.~Gaete, L.~Guajardo and S.~G.~Rodr{\'\i}guez,
``Towards the emergence of nonzero thermodynamical quantities for Lanczos-Lovelock black holes dressed with a scalar field,''
Phys. Rev. D \textbf{104} (2021) no.4, 044027
[arXiv:2103.15634 [gr-qc]].

\bibitem{Maldacena:1997re}
J.~M.~Maldacena,
``The Large $N$ limit of superconformal field theories and supergravity,''
Adv. Theor. Math. Phys. \textbf{2} (1998), 231-252
[arXiv:hep-th/9711200 [hep-th]].

\bibitem{Witten:1998qj}
E.~Witten,
``Anti de Sitter space and holography,''
Adv. Theor. Math. Phys. \textbf{2} (1998), 253-291
[arXiv:hep-th/9802150 [hep-th]].

\bibitem{GubserAdSCFT2002}
S.~S.~Gubser, I.~R.~Klebanov, and A.~M.~Polyakov, 
``A Semiclassical limit of the gauge / string correspondence,''
 Nucl. Phys. B, vol. 636, pp. 99–114, 2002.


\bibitem{Banados:2005rz}
M.~Banados, R.~Olea and S.~Theisen,
``Counterterms and dual holographic anomalies in CS gravity,''
JHEP \textbf{10} (2005), 067
[arXiv:hep-th/0509179 [hep-th]].
 \bibitem{Cvetkovic:2017fxa}
B.~Cvetkovi{\'c}, O.~Miskovic and D.~Simi{\'c},
``Holography in Lovelock Chern-Simons AdS Gravity,''
Phys. Rev. D \textbf{96} (2017) no.4, 044027
[arXiv:1705.04522 [hep-th]].

\bibitem{Banados:2006fe}
M.~Banados, O.~Miskovic and S.~Theisen,
``Holographic currents in first order gravity and finite Fefferman-Graham expansions,''
JHEP \textbf{06} (2006), 025
[arXiv:hep-th/0604148 [hep-th]].
 
\bibitem{Gallegos:2020otk}
A.~D.~Gallegos and U.~G\"ursoy,
``Holographic spin liquids and Lovelock Chern-Simons gravity,''
JHEP \textbf{11} (2020), 151
[arXiv:2004.05148 [hep-th]].

 \bibitem{Juricic:2024tbe}
V.~Juri{\v{c}}i{\'c}, O.~Miskovic and F.~R.~Carrasco,
``Holography of dislocations and ring defects in Einstein-Gauss-Bonnet AdS gravity,''
Eur. Phys. J. C \textbf{85} (2025) no.10, 1134
[arXiv:2410.18853 [hep-th]].
 


\bibitem{Regge-Teitelboim}
T.~Regge and C.~Teitelboim, 
``Role of Surface Integrals in the Hamiltonian Formulation of General Relativity,'' 
Annals Phys.\textbf{88} (1974) 286.

\bibitem{Crisostomo:2000bb} J.~Crisostomo, R.~Troncoso and J.~Zanelli,
``Black hole scan,'' Phys. Rev. D \textbf{62} (2000), 084013 [arXiv:hep-th/0003271 [hep-th]].

\bibitem{Aros:2000ij}
R.~Aros, R.~Troncoso and J.~Zanelli,
``Black holes with topologically nontrivial AdS asymptotics,''
Phys. Rev. D \textbf{63} (2001), 084015
[arXiv:hep-th/0011097 [hep-th]].

\bibitem{Palais:1979rca} R.~S.~Palais, 
``The principle of symmetric criticality,'' 
Commun.\ Math.\ Phys.\ \textbf{69}, no. 1, 19 (1979).

\bibitem{Deser:2003up} S.~Deser and B.~Tekin, ``Shortcuts to
high symmetry solutions in gravitational theories,''
Class.\ Quant.\ Grav.\ \textbf{20}, 4877 (2003) [gr-qc/0306114].



\bibitem{Deser:2004yh} S.~Deser, J.~Franklin and B.~Tekin, `
`Shortcuts to spherically symmetric solutions: A Cautionary
note,'' Class.\ Quant.\ Grav.\ \textbf{21}, 5295 (2004) [gr-qc/0404120].

\bibitem{Emparan:1999pm}
R.~Emparan, C.~V.~Johnson and R.~C.~Myers,
``Surface terms as counterterms in the AdS / CFT correspondence,''
Phys. Rev. D \textbf{60} (1999), 104001
[arXiv:hep-th/9903238 [hep-th]].


\bibitem{Mora:2004rx}
P.~Mora, R.~Olea, R.~Troncoso and J.~Zanelli,
``Vacuum energy in odd-dimensional AdS gravity,''
[arXiv:hep-th/0412046 [hep-th]].

\bibitem{Banados:2004zt}
M.~Banados, A.~Schwimmer and S.~Theisen,
``Chern-Simons gravity and holographic anomalies,''
JHEP \textbf{05} (2004), 039
[arXiv:hep-th/0404245 [hep-th]].

\bibitem{Gibbons:2004ai}
G.~W.~Gibbons, M.~J.~Perry and C.~N.~Pope,
``The First law of thermodynamics for Kerr-anti-de Sitter black holes,''
Class. Quant. Grav. \textbf{22} (2005), 1503-1526 [arXiv:hep-th/0408217 [hep-th]].

\bibitem{Banados:1998ys}
M.~Banados and F.~Mendez,
``A Note on covariant action integrals in three-dimensions,''
Phys. Rev. D \textbf{58} (1998), 104014
[arXiv:hep-th/9806065 [hep-th]].

\bibitem{Miskovic:2010ey}
O.~Miskovic and R.~Olea,
``Quantum Statistical Relation for black holes in nonlinear electrodynamics coupled to Einstein-Gauss-Bonnet AdS gravity,''
Phys. Rev. D \textbf{83} (2011), 064017
[arXiv:1012.4867 [hep-th]].

\bibitem{Regge:1974zd}
T.~Regge and C.~Teitelboim,
``Role of Surface Integrals in the Hamiltonian Formulation of General Relativity,'' 
Annals Phys. \textbf{88} (1974), 286.

\bibitem{Dordevic:2026pct}
D.~{\DJ}or{\dj}evi{\'c} and D.~Go{\v{c}}anin,
``Holographic entanglement entropy in Chern-Simons gravity with torsion,''
[arXiv:2602.12197 [hep-th]].

\bibitem{Araneda:2016iiy}
R.~Araneda, R.~Aros, O.~Miskovic and R.~Olea,
``Magnetic Mass in 4D AdS Gravity,''
Phys. Rev. D \textbf{93} (2016) no.8, 084022
[arXiv:1602.07975 [hep-th]].

\bibitem{Wald:1993nt} R.~M.~Wald, ``Black hole entropy is
the Noether charge,'' Phys. Rev. D \textbf{48} (1993) no.8,
R3427-R3431 [arXiv:gr-qc/9307038 [gr-qc]].


\bibitem{Banados:2001hm}
M.~Banados, ``Charged solutions in 5d Chern-Simons supergravity,''
Phys. Rev. D \textbf{65} (2002), 044014
[arXiv:hep-th/0109031 [hep-th]].

%\cite{Kallosh:1992ii}
\bibitem{Kallosh:1992ii}
R.~Kallosh, A.~D.~Linde, T.~Ortin, A.~W.~Peet and A.~Van Proeyen,
``Supersymmetry as a cosmic censor,''
Phys. Rev. D \textbf{46} (1992), 5278-5302
doi:10.1103/PhysRevD.46.5278
[arXiv:hep-th/9205027 [hep-th]].


\bibitem{Kallosh:1992wa}
R.~Kallosh, T.~Ortin and A.~W.~Peet,
``Entropy and action of dilaton black holes,''
Phys. Rev. D \textbf{47} (1993), 5400-5407
doi:10.1103/PhysRevD.47.5400
[arXiv:hep-th/9211015 [hep-th]].


\bibitem{Ferrara:1995ih}
S.~Ferrara, R.~Kallosh and A.~Strominger,
``N=2 extremal black holes,''
Phys. Rev. D \textbf{52} (1995), R5412-R5416
doi:10.1103/PhysRevD.52.R5412
[arXiv:hep-th/9508072 [hep-th]].

\bibitem{Ferrara:1997tw}
S.~Ferrara, G.~W.~Gibbons and R.~Kallosh,
``Black holes and critical points in moduli space,''
Nucl. Phys. B \textbf{500} (1997), 75-93
doi:10.1016/S0550-3213(97)00324-6
[arXiv:hep-th/9702103 [hep-th]].

\bibitem{Strominger:1996sh}
A.~Strominger and C.~Vafa,
``Microscopic origin of the Bekenstein-Hawking entropy,''
Phys. Lett. B \textbf{379} (1996), 99-104
doi:10.1016/0370-2693(96)00345-0
[arXiv:hep-th/9601029 [hep-th]].


\end{thebibliography}
\end{document}